\documentclass[journal,twoside]{IEEEtran}

\usepackage{cite}
\usepackage{amsmath}
\usepackage{graphicx}
\usepackage{amssymb}
\usepackage{array}
\usepackage{balance}
\usepackage{dsfont}
\usepackage{bbm}
\usepackage{ifpdf}
\usepackage[normalem]{ulem}

\usepackage[T1]{fontenc}
\graphicspath{{./Figuras/}}                   

\newtheorem{corollary}{Corollary}[section]
\newtheorem{Lemma}{Lemma}[section]
\newtheorem{Theorem}{Theorem}[section]
\newtheorem{remark}{Remark}[section]

\newcommand{\prob}{\mathbb{P}}
\newcommand{\adf}{\mathcal{A}}
\newcommand{\bdf}{\mathcal{B}}
\newcommand{\adt}{\mathcal{A}_{DT}}

\newcommand{\poutdt}{\prob_{\text{out,DT}}}

\newcommand{\Ex}{\mathbb{E}}
\newcommand{\R}{\mathbb{R}}

\newcommand{\Onot}{O}
\newcommand{\erf}{\text{erf}}
\newcommand{\poutm}{\prob_{\text{out,mix}}}
\newcommand{\bigone}{\mathds{1}}
\usepackage{color}
\definecolor{red}{rgb}{1,0.5,0}

\begin{document}
\title{Analysis of a Cooperative Strategy for a Large Decentralized Wireless Network}

\author{Andr\'es Altieri,
        Leonardo Rey Vega,
        Pablo Piantanida
        and Cecilia G. Galarza
\thanks{This work was partially supported by DIGITEO-DIM  No. 2010-33D (ACRON), the Peruilh grant of the UBA and project UBACYT 2002010200250. The material in this paper was presented in part at the IEEE International Symposium on Information Theory, 2011 and 2012.}
\thanks{L. Rey Vega and C. Galarza are with the Department of Electronics (FIUBA) and CONICET, Buenos Aires, Argentina (e-mail: lrey@fi.uba.ar, cgalar@fi.uba.ar).}
\thanks{P. Piantanida is with the Department of Telecommunications, SUPELEC, 91192 Gif-sur-Yvette, France (e-mail: pablo.piantanida@supelec.fr).}
\thanks{A. Altieri (e-mail: aaltieri@fi.uba.ar, andres.altieri@supelec.fr).}
}

%


\maketitle

\begin{abstract}
This paper investigates the benefits of cooperation and proposes a relay activation strategy for a large wireless network with multiple transmitters. In this framework, some nodes cooperate with a nearby node that acts as a relay, using the decode-and-forward protocol, and others use direct transmission. The network is modeled as an independently marked Poisson point process and the source nodes may choose their relays from the set of inactive nodes. Although cooperation can potentially lead to significant improvements in the performance of a communication pair, relaying causes additional interference in the network, increasing the average noise that other nodes see. We investigate how source nodes should balance cooperation vs. interference to obtain reliable transmissions, and for this purpose we study and optimize a relay activation strategy with respect to the outage probability. Surprisingly, in the high reliability regime, the optimized strategy consists on the activation of all the relays or none at all, depending on network parameters. We provide a simple closed-form expression that indicates when the relays should be active, and we introduce closed form expressions that quantify the performance gains of this scheme with respect to a network that only uses direct transmission.
\end{abstract}

\begin{IEEEkeywords}
Cooperative communication, interference, network management, outage probability, decode and forward, marked Poisson point processes. 
\end{IEEEkeywords}

%
\IEEEpeerreviewmaketitle

\section{Introduction}
\label{sec:intro}

Cooperative wireless networks in which relay nodes can be used to increase throughput and reliability have been studied in the past \cite{kramer_cooperative_2005}. Although the capacity of the single-relay channel \cite{cover_capacity_1979} remains unsolved and its optimal coding scheme unknown, there have been significant advances in quantifying the performance gain obtained through cooperation. However, finding capacity regions or analyzing the performance of large random wireless networks may be, if feasible, very hard. As an alternative, spatial models employing tools from stochastic geometry and graph theory provide a comprehensive framework for the analysis of large wireless networks with little interference management \cite{baccelli_aloha_2006, weber_overview_2010}. 

The \emph{outage probability} (OP) and its complement, the \emph{success probability}, are useful metrics in decentralized networks \cite{baccelli_aloha_2006}, \cite{weber_overview_2010}, \cite{HGG2011} in which the users are assumed to be unaware of the instantaneous parameters of the network and cannot optimize their behavior to attain successful transmissions. Among other reasons, the relevance of the OP comes from the fact that, in an outage event, sent messages cannot be successfully transmitted, and hence the overall delay of the network is increased due to retransmissions. In this paper we investigate the performance, in terms of OP, of a large decentralized wireless network in which transmitters may be aided by nearby relays. More precisely, we consider a network formed by two types of clusters:  source-relay-destination clusters, which use the full-duplex \emph{decode-and-forward} (DF) \cite{cover_capacity_1979} scheme, and clusters with source-destination pairs which employ simple \emph{direct transmission} (DT). These clusters could be interpreted as a single hop in a multi-hop transmission scheme or by themselves as single-hop communications.
One of the central motivations behind this analysis is to provide an understanding of the limitations and benefits of cooperation in such decentralized scenarios. In fact, the advantage of cooperation among nodes for an individual source-destination link was widely studied in the past years, addressing both theoretical and practical issues \cite{kramer_cooperative_2005, laneman_cooperative_2004, host-madsen_capacity_2005}.  {In this paper we analyze a scenario in which the communication impairments are caused by a network of users which are also attempting to achieve successful transmissions through cooperation and cause interference to each other.} It is clear that relays can significantly improve the rate and reliability of a single source-destination pair. However, in a large wireless network, the nodes will observe an increase in their interference levels as more relays are activated. This means that while cooperation may be beneficial locally, globally its benefits may be reduced. 

In this paper, the network is modeled as an independently marked homogeneous {\it Poisson point process} (PPP) \cite{kingman93}, limited by the \emph{signal-to-interference ratio} (SIR), where signal attenuation occurs both through path loss and slow fading \cite{GoldsmithWC}. The random distribution of the sources and their relays implies that, in addition to the random fading, averaging over all spatial positions is needed to derive the OP. We focus on the high reliability regime as defined in \cite{HGG2011}.
{In that work, the authors study the outage behavior of general motion-invariant networks employing DT, by resorting to an asymptotic analysis in which the density of interfering nodes goes to zero. In particular they show that the OP using an arbitrary medium access scheme approaches $ 1 - \gamma \lambda_s^\kappa$ as the density of interferers $\lambda_s \rightarrow 0$, where $\gamma$ is the \emph{spatial contention} parameter and $\kappa$ is the \emph{interference scaling exponent}. For the case of networks using the ALOHA access scheme we have that $\kappa = 1$. The \emph{high reliability regime} as defined in \cite{HGG2011} refers to the operating regime in which the OP is small enough (close to zero) to guarantee that the asymptotic first order approximation is a good representation of the network performance. This regime covers OP values of the order of $0.01$ which are typical in wireless system designs \cite{GoldsmithWC}.}
An outage event is declared whenever the distribution of nodes and/or fading cause the chosen rate to be higher than the achievable rate of the transmission protocol of choice. Hence, the probability of these events (OP) is an upper bound on the asymptotic packet error probability of every pair of communicating nodes, which is a key metric of interest \cite{altieri_cooperative_2012}.

\begin{figure}[!t]
\centering
\ifpdf	\includegraphics[angle=0,width=.9\columnwidth,keepaspectratio,trim= 0 0 0 0,clip]{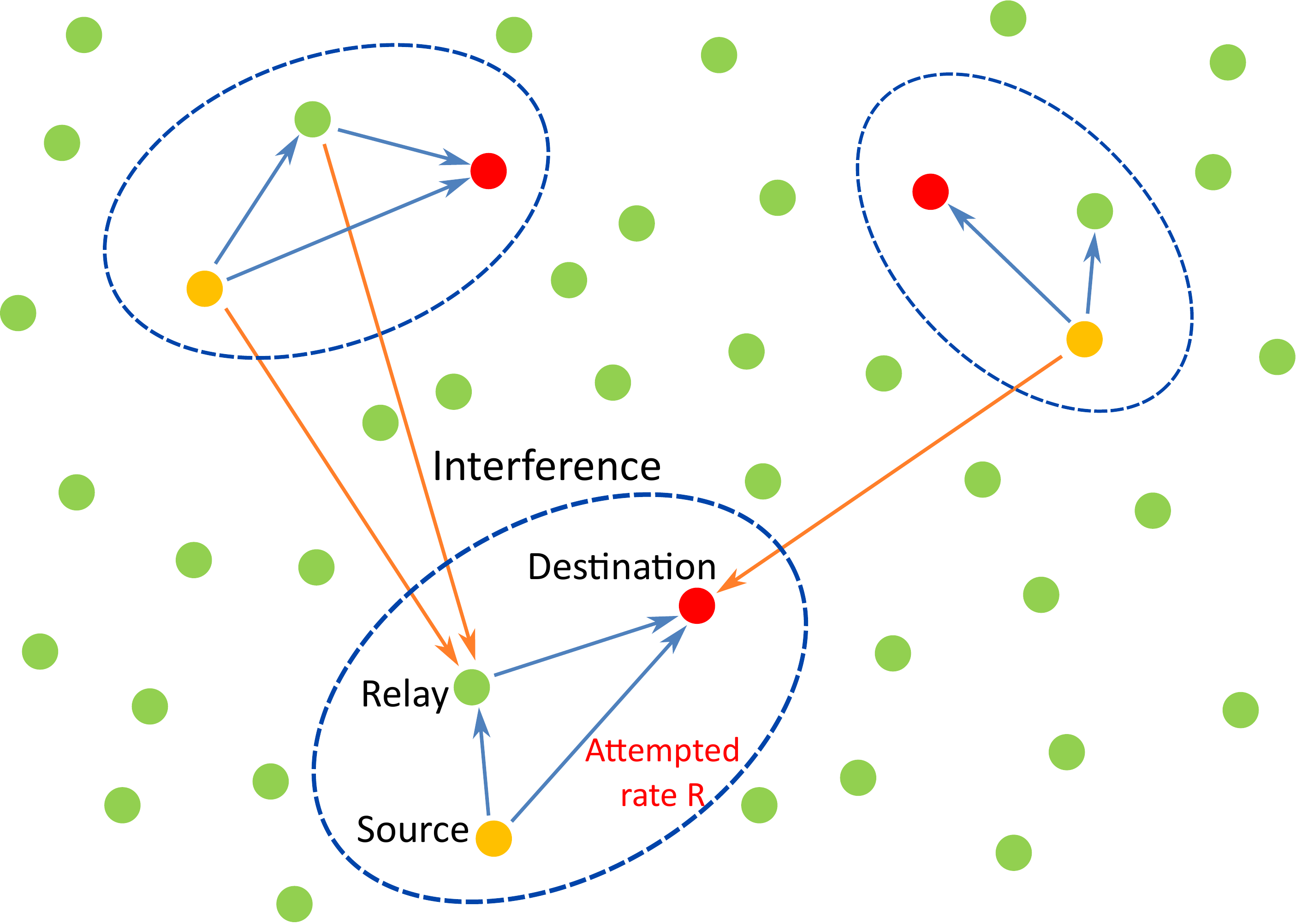} 
\else
	\includegraphics[angle=0,width=.9\columnwidth,keepaspectratio,trim= 0 0 0 0,clip]{relays.eps} 
\fi
\caption{The network is formed of clusters employing decode-and-forward or direct transmission.}
\label{fig:relay}
\end{figure}

The transmission scheme of the network is a mixed one, since some clusters will be using the DF scheme while others will employ DT (see Fig. \ref{fig:relay}). It is assumed that almost no \emph{channel-state information} (CSI) is available at the transmitting nodes, which is often the case in decentralized wireless networks without feedback. Only a rough estimation of the position of nearby potential relays may be available, and hence, it can be used for the relay selection. We assume that each source chooses its potential relay among the nodes that are not transmitting as its \emph{nearest neighbor} (NN) on a cone with aperture angle $\phi_0$, centered toward its destination (see Fig. \ref{fig:cone}). This scheme will increase the likelihood of finding a relay which is close to the source and at the same time reduces the effect of the path loss on the relay-destination link. Notice that this effect is minimized if the relay, source and destination are aligned. As a special case of this scenario, the relay can be chosen as the NN of the source on the whole plane, requiring the least amount of CSI. The motivation behind choosing the NN as a relay comes from the fact that decode-and-forward is nearly optimal from an information theoretic point of view \cite{kramer_cooperative_2005} \cite{cover_capacity_1979}  \cite{altieri_cooperative_2011} when the relay is not far from the source. In this case, the probability of the relay not being able to decode the source message is minimized. A simple random relay activation scheme is introduced in which each candidate relay node decides whether to be active or not in a random manner, independently of each other, and of all network parameters. This simple activation scheme will act as a means of controlling the relay density in the network while still retaining a balance between interference generation and cooperation.

\subsection{Related Work}
\label{subsec:related}
Over the past years, the performance gains of cooperative communications in relay networks were widely studied from an information-theoretic perspective. Since the seminal work of Cover and El Gamal \cite{cover_capacity_1979}, several contributions have been published on the subject. More recently, the emphasis has been put on studying the performance of wireless relay channels where outage performance and ergodic rates of fading channels with Gaussian noise have been derived (see \cite{kramer_cooperative_2005,laneman_cooperative_2004,host-madsen_capacity_2005,katz_cooperative_2009} and the references therein). Among these valuable studies, the only impairments to the communication were due to additive Gaussian noise and fading, and very little attention was paid to the effect of the interference generated (or suffered) by other users. However, interference is probably the major impairment in wireless networks, specially in networks with little control and high mobility.

The study of the capacity of general wireless networks taking into account the interference generated by the different users was pioneered by the seminal work of Gupta-Kumar \cite{gupta_capacity_2000}, where the concept of transport capacity and fundamental scaling laws on the network throughput were obtained considering only point-to-point coding. In \cite{gupta_towards_2003} multiuser achievability regions were obtained and it was shown that for some special wireless networks significantly better scaling laws on the network throughput, with respect to the case in \cite{gupta_capacity_2000},  are possible. Further progress was done in \cite{liang-liang_xie_network_2004}, where new scaling laws were derived using coherent multistage relaying with interference subtraction and in \cite{xue_transport_2005}, where extensions to fading channels were obtained.


Stochastic geometry and point processes \cite{BB2010,DVJ1998} are not only elegant mathematical frameworks but also useful tools to deal with more realistic network models, where the spatial position of nodes and the effect of interference can be incorporated in a probabilistic manner \cite{baccelli_aloha_2006}. Although several types of point processes can be used to model different kind of networks, it is the homogeneous PPP which has received the most	attention. Although other types of point processes could provide more realistic models \cite{ganti_interference_2009},  the extended use of the homogeneous PPP comes from the possibility to obtain simple closed-form results in several cases of interest. The quantity called transmission capacity (TC) was introduced in \cite{weber_transmission_2005} in order to include outage probability constraints in the scaling behavior. Several results have been obtained, through the use of the TC, for several practical situations, as multiple input-multiple output capable users in wireless networks \cite{hunter_transmission_2008}, decentralized power control \cite{jindal_fractional_2008}, etc. (for a review of several other important results please see \cite{weber_overview_2010} and the	 references therein).   
\subsection{Main Contributions}
\label{subsec:contributions}

The main contribution of this paper is studying and optimizing the network management strategy for activating the relays in the proposed transmission scheme. The main conclusion is that in the high reliability regime this optimized scheme consists on turning all the relays on or off simultaneously, that is, the optimal relay activation probability is either $0$ or $1$. To do this, we develop closed-form approximations for the OP of the network, and study the interference-cooperation balance by finding the relay activation probability which minimizes the average OP. Moreover, the network parameter regions in which all the relays should be on are identified, and a simple relay activation scheme which is close to the optimal behavior is introduced. Finally we provide simple expressions that quantify the performance gains in terms of OP for the scheme with the optimal relay activation scheme with respect to a network in which all users employ DT.

The paper is organized as follows. In Section \ref{sec:model}, a general and a mathematical descriptions of the network model are presented. We also discuss the DF scheme and its achievable rate in the assumed network model. In Section \ref{sec:OP} we introduce an expression for the OP for this network, deriving closed form approximations to it. In Section \ref{subsec:randomrelays} we study the performance of the network, finding the optimal relay activation probability, identifying the network parameters for which all the relays should be on or off, introducing the relay activation policy and comparing the performance of this scheme against DT. In Section \ref{sec:numerical} we present some numerical simulations and in Section \ref{sec:final} we provide some concluding remarks. Finally, long mathematical proofs are grouped together by section and deferred to the appendices.

\subsection*{Notation} $\mathbb{R}$, $\mathbb{C}$, $\mathbb{R}^2$ and $\|\cdot\|$ denote the real and complex numbers, the real plane and the canonical euclidean norm, respectively. $(\cdot)^{*}$ and $\Re(\cdot)$  denote complex conjugation and the real part of complex number. $\Ex_X\left[\cdot\right]$ denotes expectation with respect to the random variable $X$.
We shall use the big O notation: $f(x)=\Onot\left(g(x)\right)$ as $x\rightarrow x_0$ if there exists $M>0$ and such that $|f(x)|\leq M|g(x)|$ is some neighborhood of $x_0$. Finally, $\bigone(x\in A)$ denotes the indicator function, which is $1$ if $x\in A$ and $0$ otherwise.

\begin{figure}[!t]
\centering
\ifpdf
\includegraphics[angle=0,width=0.9\columnwidth,keepaspectratio,trim= 0 0 0 0,clip]{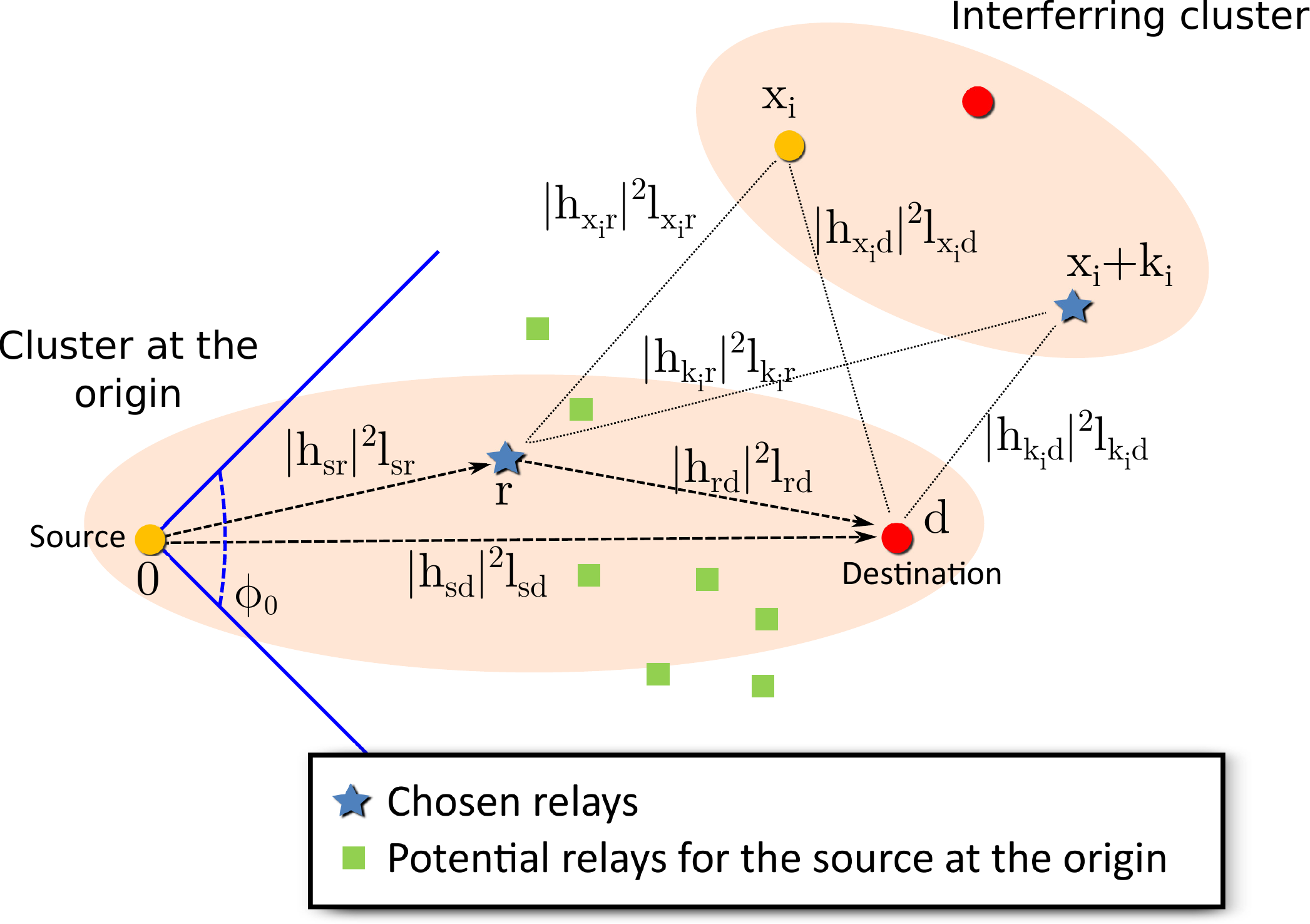} 
\else
	\includegraphics[angle=0,width=.9\columnwidth,keepaspectratio,trim= 0 0 0 0,clip]{cone2.eps} 
\fi
\caption{The relay is chosen as the nearest neighbor of the source on a cone of aperture $\phi_0$ with its axis aligned with the destination. Also the power fading within the cluster at the origin and for the interference from other clusters are shown.}
\label{fig:cone}
\end{figure}
\section{General considerations and network model}
\label{sec:model}
\subsection{The Model}
We consider a spatial network model in $\mathbb{R}^2$ in which source nodes generate messages and attempt to transmit them to intended destinations, either through a direct link, in which case the destinations receive symbols from their sources only, or by using others nodes as relays. Every relay aids a single source node, acting as a secondary full-duplex transmitter sharing the same time slots and frequency band. This setup allows the nodes to be grouped into \emph{clusters} formed by a source-destination pair or by a source-relay-destination triplet, if the source has an associated relay, as shown in Fig. \ref{fig:relay}.

We start from a set of nodes $\Phi$ which we assume forms an homogeneous PPP of density $\lambda$. Some nodes from this set choose to access the network and become sources using slotted ALOHA \cite{baccelli_aloha_2006} with transmit probability $\lambda_s / \lambda$. This splits the set $\Phi$ into two new independent homogeneous PPPs:
\begin{itemize}
\item $\Phi_s$ of sources of density $\lambda_s$,
\item $\Phi_{in}$ of  potential relay nodes of density $\lambda_{in}=\lambda-\lambda_{s}$,
\end{itemize}
such that $\Phi = \Phi_s \cup \Phi_{in}$. Notice that the proportion of sources and potential relays can be adjusted by the medium access probability. 

Inactive nodes should then be assigned in a one-to-one fashion to each source such that cooperation is beneficial. To simplify the relay assignment strategy we shall assume that the spatial density of the sources is much smaller than that of potential relays, i.e. $\lambda_s \ll \lambda_{in}$. Under this hypothesis, we will neglect the probability of two sources choosing the same inactive node as a relay, since each source will have a rich selection of relay candidates in its vicinity (\cite{WAJ2007}, Ex. 3). Thus we can simplify our model by including the position of the potential relay and its activation scheme as an independent mark to each source, obtaining the spatial distribution of the relay from the original homogeneous PPP $\Phi_{in}$ of intensity $\lambda_{in}$ to which the relays are assumed to belong. 

We consider the usual and realistic assumption that only little or no CSI is available, while nodes may have some estimation of the spatial position of neighboring nodes. For this reason nodes cannot adjust their rates to achieve a reliable communication according to instantaneous conditions, but may use this spatial knowledge to select a relay.

Based on these considerations, the network is modeled as an independently marked PPP:
\begin{equation}
\tilde{\Phi}_s= \left\{ (x_i,(\varepsilon_{x_i},k_i, \theta_{x_i}),h_{x_ir},h_{x_id},h_{k_ir},h_{k_id},)\right\},
\end{equation}
such that:
\begin{itemize}
\item The positions of the sources form the homogeneous PPP $\Phi_s = \left\{ x_i \right\}$ of intensity $\lambda_s$. 
\item The triplet $(\varepsilon_{x_i},k_i, \theta_{x_i})$ models the relay position and its state. The random variable (RV) $\theta_{x_i}$, uniform in $[0,2\pi)$, models the direction of each destination relative to its source, with $\theta= 0$ meaning that the destination is in the direction of the canonical vector  $(1,0)$ with respect to its source. $k_i$ indicates the position of the potential relay relative to its source, that is, the potential relay for source $x_i$ is located at $x_i+k_i$. According to what we mentioned earlier, the relay will be chosen as the NN of the source on a cone of aperture $\phi_0$ with the destination on its axis. This means that the distribution of the potential relay $k_i$ for a source at the origin, conditioned on the direction of the destination $\theta_{x_i}$ will be (in polar coordinates) \cite{H2005routing}:
\begin{multline}
f_{k_i|\theta} (\rho, \phi) = \lambda_{in} \rho e^{-\lambda_{in} \phi_0 \rho^2/2} \bigone\{|\phi-\theta| < \phi_0/2\} \\ \times \bigone\{0\leq \phi \leq 2\pi, \rho \geq 0\}. \label{eq:NNdist1}
\end{multline}

Using $\phi_0 = 2\pi$ means choosing the relay as the NN on the whole plane instead of a cone, independently of the direction of the destination and using the least CSI. Notice that in this case the distribution of the NN (\ref{eq:NNdist1}) becomes a bidimensional Gaussian RV of variance:
\begin{equation}
\sigma_{in}^2 = \frac{1}{{2 \pi \lambda_{in}}}.
\end{equation}
Notice that we can parameterize the NN distribution (\ref{eq:NNdist1}) in terms of $\sigma_{in}$ and for any cone aperture $\phi_0$ as:
\begin{multline}
f_{k_i|\theta} (\rho, \phi) = \frac{1}{2\sigma_{in}^2} \rho e^{- \frac{\phi_0 \rho^2}{4\pi \sigma_{in}^2}} \bigone\{|\phi-\theta| < \phi_0/2\} \\ \times \bigone\{0\leq \phi \leq 2\pi, \rho \geq 0\}.
\end{multline}
This implies that the effect of considering the nearest neighbor on a cone is simply restricting the NN on the plane distribution (Gaussian) to the cone and increasing the variance (by means of the $\phi_0$ in the exponent). Thus, we can study the relay activation strategy in terms of the variance of the nearest neighbor on the whole plane ($\sigma_{in}$) and cone aperture $\phi_0$ that the source uses. Additionally, notice that reducing the cone aperture allows the relay to be located towards the destination but at the same time, the increased variance implies that the relay will be, on average, farther from the source than if we take $\phi_0 = 2\pi$.

The RV $\varepsilon_{x_i}$ indicates if the corresponding source uses a relay or not. In our case, we take it to be a \emph{Bernoulli} RV with success probability $p_r$, independent of everything else.
Notice that the parameter $p_r$ allows the adjustment of the relay density and hence allows to control the additional interference introduced in the network, weighing the local and global effects of cooperation. In addition to a MAC access scheme, this parameter can also be used to model the unavailability of a relay for reasons which are out of the control of the relay itself (such as a malfunction or a depleted battery). In such a case, the independent occurrence of these events among the relays is a reasonable assumption.
\item All nodes transmit with unit power while the power received at $y$ by a transmitter located at $x$ is $|h_{xy}|^2
    l_{xy}$ where $l_{xy}= \|x-y\|^{-\alpha}$ ($\alpha >2$) is the usual path loss function and $|h_{xy}|^2$ is the power gain of Rayleigh fading 
    with unit mean. This is equivalent to saying that $h_{xy}$ are complex, circular  \cite{picinbono_circularity_1994}, zero-mean Gaussian RVs.
\item An additional source with the same marks as the others, independent of the point process $\tilde{\Phi}_s$ and with its destination at $d = (D,0)$, is added at the origin. The position of the relay for this source node will be $r$ (with the same distribution as the $\{k_i\}$ RVs). The coefficients $|h_{sr}|^2$, $|h_{rd}|^2$ and $|h_{sd}|^2$ model the source-relay, relay-destination, and source-destination fading coefficients of this cluster, respectively.  
Slyvniak's Theorem \cite{BB2010}, \cite{DVJ1998} guarantees that the study of this cluster's behavior will be representative of the behavior of any other similar cluster in the network and hence it can be considered as a \textquotedblleft typical cluster''.
\item $h_{x_ir}$ and $h_{k_ir}$ model the fading gains from each source and its relay to the relay of the source at the origin, while $h_{x_id}$, $h_{k_id}$ model the gains from each source and its relay to the destination of the source at the origin. 
\end{itemize}\vspace{2mm}

Please see Fig. \ref{fig:cone} for a graphical representation of the key parameters in the model.
\begin{remark}
Notice that other schemes for activating and selecting the relays based on position can be studied by appropriately selecting the triplet $(\varepsilon_{x_i},k_i, \theta_{x_i})$. For example, we could study the performance of choosing the nearest or the farthest neighbor on a cone of finite radius as a relay. 
The probability of activating a relay will be that of finding at least one potential relay in the cone and the conditional distributions of the selected relay, given that the cone is not empty, can be found in \cite{H2005distances}. 
\end{remark}

It is assumed that during the transmission time all the positions of the nodes, fading coefficients and other network parameters encompassed in the marked PPP $\tilde{\Phi}_s$  remain constant, that is, there is no node mobility and a slow fading scenario is considered. Within a cluster, each source and its relay (if it is active) use Gaussian signaling and their codebooks have correlation coefficient $\rho$. In addition, the codebooks between different clusters are independent. Destination and relay nodes in each cluster attempt to decode their messages while treating the interference from other clusters as noise. With these hypotheses  the following Lemma can be proved:
\begin{Lemma} \label{lemma:tightness}
If $\alpha>2$, then for almost all realizations of the point process $\tilde{\Phi}_s$, the aggregate interferences at the relay and destination of the typical cluster are zero-mean complex circular Gaussian variables whose variances conditioned on $\tilde{\Phi}_s$ are given by:
\begin{multline}
I_r = \sum_{i:x_i\in\Phi_s}\left[\frac{|h_{x_ir}|^2}{\|x_i-r\|^{\alpha}}\right.\\ \left.
+\varepsilon_{x_i}\!\left(\!\frac{|h_{k_ir}|^2}{\|x_i+k_i-r\|^{\alpha}}\!
\hspace{-1pt} +\!\frac{2\Re\left\{h_{x_ir}h_{k_ir}^*\rho\right\}}{\|x_i-r\|^{\frac{\alpha}{2}}\|x_i+k_i-r\|^{\frac{\alpha}{2}}}\right)\!\right]\hspace{-2pt}, \hspace{-6pt}
\label{eq:power_inter_relay}
\end{multline}
\vspace{-3mm}
 \begin{multline}
I_d = \sum_{i:x_i\in\Phi_s}\left[\!\frac{|h_{x_id}|^2}{\|x_i-d\|^{\alpha}}\right.
\left.\!\right.\\\left.
+\varepsilon_{x_i}\!\left(\!\frac{|h_{k_id}|^2}{\|x_i+k_i-d\|^{\alpha}}\!
+\!\frac{2\Re\left\{h_{x_id}h_{k_id}^*\rho\right\}}{\|x_i-d\|^{\frac{\alpha}{2}}\|x_i+k_i-d\|^{\frac{\alpha}{2}}}\right)\!\right] \hspace{-2pt}. \hspace{-10pt}
\label{eq:power_inter_dest}
\end{multline}
\end{Lemma}
\begin{IEEEproof} 
See appendix \ref{proof:tightness}.
\end{IEEEproof}

\subsection{Achievable Rates} 

The main coding strategies for relay networks were introduced in the seminal work by Cover and El Gamal \cite{cover_capacity_1979}. There have been three dominant relaying paradigms: decode-and-forward (DF), compress-and-forward (CF), and amplify-and-forward (AF). In AF, the relay simply repeats an amplified version of what it received without decoding the message. In CF, the relay chooses an appropriate sequence from a set that acts as a compressed version of what it received. In DF, the relay decodes the messages sent by the source, re-encodes it, and forwards it to the destination, which decodes the message by using both the transmission from the source and from the relay.
In general, DF will work best when the source-relay channel is good enough to avoid a bottleneck in the information flow with respect to a source-destination transmission. In a scenario in which the spatial distribution of nodes is considered, the quality of the source-destination and source-relay channels will be heavily influenced, through the path loss, by the distances between the nodes. This means that in order to avoid this bottleneck, the relay should be chosen so that on average the source-relay distance is smaller than the source-destination distance.
Other variations of DF such as partial-decode-and-forward  \cite{cover_capacity_1979} relax the imposition of full-decoding at the relay; however, they require a careful optimization of the code at the encoder, which cannot be done in our setting due to the lack of CSI at the source.

There are several encoding and decoding techniques which implement the DF scheme, all of which are based on block-Markov encoding. For our analysis of the error events we consider either \emph{regular encoding} and \emph{sliding-window decoding}  \cite{carleial_multiple-access_1982} at the destination or \emph{regular encoding} and \emph{backward decoding} \cite{willems_discrete_1985,zeng_achievability_1989}\footnote{Another alternative using   \emph{irregular encoding}, \emph{random binning} and \emph{successive decoding} at the destination was introduced in \cite{cover_capacity_1979} but it is not suited for our analysis since additional error events have to be considered.}. With any of these two schemes, conditioned on a particular realization of $\tilde{\Phi}_s$, the relay channel associated to the source located at the origin can achieve a rate \cite{cover_capacity_1979}:	
\begin{multline}
R_{DF} = \max_{\rho\in\mathbb{C}, |\rho|\leq 1}\min \left\{  \mathcal{C} \left(\frac{|h_{sr}|^2l_{sr}(1-|\rho|^2)}{I_r}\right),
\right.\\ \left.
\mathcal{C}\left(\frac{|h_{sd}|^2l_{sd}+|h_{rd}|^2l_{rd}+2\sqrt{l_{sd}l_{rd}}\Re\left(\rho
h_{sd}h_{rd}^{*}\right)}{I_d}\right) \right\}, \label{eq:DF_rate_gaussian_poisson}
\end{multline}
where $\mathcal{C}(u) = \log_2(1+u)$. The maximization with respect to $\rho$ is considered because the correlation of the codebooks within a cluster affects the interference seen by other nodes and also the achievable rates for each cluster. This means that in general the value of $\rho$ that maximizes the achievable rate could be selected \cite{kramer_cooperative_2005}\cite{cover_capacity_1979}. In this work, we shall consider the case $\rho = 0$, which simplifies the implementation of DF,  as pointed out in \cite{kramer_cooperative_2005} (see remark 42) and \cite{host-madsen_capacity_2005},  because symbol synchronization between the source and its corresponding relay, is not strictly required. Although other choices of $\rho$ could improve the outage behavior of the network, $\rho = 0$ is known to be the optimal value in the high reliability regime for a network in which only one source is allowed to use a relay \cite{altieri_cooperative_2011}. Therefore, when the relay is present we can define the outage event $\adf(R)\cup \bdf(R) $ as
\begin{IEEEeqnarray*}{rCl}
\adf(R)&=&\left\{|h_{sr}|^2 l_{sr}<T I_r \right\}, \\
\bdf(R) &=& \left\{ |h_{sd}|^2 l_{sd} + |h_{rd}|^2 l_{rd} < TI_d\right\},
\end{IEEEeqnarray*}
where $R$ is the attempted rate by the source and $T= 2^R-1$. The event $\adf(R)$ means that the relay is in outage while $\bdf(R)$ means that the destination is in outage while source and relay cooperate. 


The DF scheme with backward or sliding-window decoding at the destination are \emph{oblivious} \cite{katz_cooperative_2009} to the presence of the relay, that is, the source can use the same coding scheme for DF or DT without considering if the relay is present or not. This is very important, since the relay can decide to activate itself (achieving the DF rate) or not (achieving the DT rate) without taking into account the source, which in both cases employs the same coding scheme. Only the destination knows if the relay is present and can adapt its decoding strategy suitably according to each case. Also, the rate $R_{DF}$ does not depend on the correlation between the noises or interferences at the relay and the destination. This is true because the correlation between received signals at the relay and the destination becomes irrelevant when full decoding at the relay is imposed. As a matter of fact, this is not the case for the CF and AF schemes where the correlation between the noises can increase or decrease the corresponding achievable rate \cite{lili_zhang_study_2011}.

Finally, we also define the outage event for the case in which there is no relay and thus the source simply uses DT \cite{baccelli_aloha_2006}:
\begin{equation}
\adt(R) = \left\{ \frac{|h_{sd}|^2 l_{sd}}{I_d} < T\right\}. \label{eq:adt}
\end{equation}
The probability of this event is known to be\cite{baccelli_aloha_2006}
$\poutdt(R) = 1 - e^{-\lambda_s \delta D^2}$,
where:
\vspace{-3mm}
\begin{gather}
\delta = C T^{2/\alpha}, \\
C = \frac{2\pi }{\alpha} \Gamma\left(\frac{2}{\alpha}\right)  \Gamma\left(1-\frac{2}{\alpha}\right), \label{eq:C}
\end{gather}
and $\Gamma(u) = \int_0^\infty t^{u-1} e^{-t} dt$ is the Gamma function. Using the asymptotic expansion of the OP we can write $\poutdt(R) = \lambda_s \delta D^2 + \Onot((\lambda_s \delta D^2)^2)$ as $(\lambda_s \delta D^2)^2 \rightarrow 0$. In the high reliability regime, when the success probability of the network is close to one, a reasonable approximation is to neglect the higher order $O(\cdot)$ term and write $\poutdt(R) \approx \lambda_s \delta D^2$, meaning that the approximation will be good and that $\lambda_s \delta D^2$ will be small. In this expansion we see that $\gamma = \delta D^2$ is the contention parameter of the network, as defined in Section I.

\section{The outage probability of the network} \label{sec:OP}

In this section we study the OP of the network as introduced in the previous section. By conditioning on the fact that the cluster at the origin uses a relay or not, and on this relay position, we can see that the OP of the cluster at the origin (and hence of any other cluster) can be written as:
\begin{multline}
\poutm (R) =  \prob\{\varepsilon_0 = 0\}\prob\left\{\adt(R) | \varepsilon_0 = 0\right\} \\
+ \prob\{\varepsilon_0 = 1\} \Ex_r \left[ \prob\left\{\adf(R) \cup \bdf(R)  |r, \varepsilon_0 = 1 \right\}\right]. \label{eq:genop}
\end{multline}
This expression can be evaluated in terms of the Laplace transform of the interference random variables $I_r$ and $I_d$, as the following theorem states:
\begin{Theorem} \label{teo:opexp}
The outage probability of the network $\poutm$ given by (\ref{eq:genop}) can be written as:
\begin{multline}
\poutm (R) = \prob\{\varepsilon_0 = 0\} \left[1 - \mathcal{L}_{I_d} \left(T /l_{sd}\right)\right] + \prob\{\varepsilon_0 = 1\}\\
\times \hspace{-1pt}\Ex_r \hspace{-3pt}\left[ \frac{D^\alpha \mathcal{L}_{I_d,I_r} \hspace{-2pt} \left(\frac{T}{l_{rd}},\frac{T}{l_{sr}}\right) \hspace{-2pt} - \hspace{-2pt} ||r-d||^\alpha \mathcal{L}_{I_d,I_r} \hspace{-2pt} \left(\frac{T}{l_{sd}},\frac{T}{l_{sr}} \right)}{D^\alpha - ||r-d||^\alpha}\right] \hspace{-10pt}
\end{multline}
where:
\begin{equation}
\mathcal{L}_{I_d,I_r}\left( \omega_1, \omega_2\right) := \Ex_{\tilde{\Phi}_s} \left[ e^{-(\omega_1 I_d + \omega_2 I_r)} \right],\ \omega_1,\omega_2\in\mathbb{C},
\label{eq:lapdef}
\end{equation}
with $\Re\left\{\omega_1\right\}, \Re\left\{\omega_1\right\}>0 $ is the joint Laplace transform of the interference at the relay and at the destination. Additionally, setting $\omega_2 = 0$ in (\ref{eq:lapdef}) we obtain $\mathcal{L}_{I_d}(\omega_1)$, the Laplace transform of the interference at the destination.
\end{Theorem}
\begin{IEEEproof}
See appendix \ref{proof:opexp}.
\end{IEEEproof}
The Laplace transforms of interference RVs are known in closed form in some special cases only, and in general they can only be expressed in terms of integrals in several dimensions (see \cite{ABG2012} and the references therein). A brief review on them can be found in appendix \ref{ap:Laptrans}. Now, using (\ref{eq:Lapgen1}) the two-dimensional Laplace transform $\mathcal{L}_{I_d,I_r} (\omega_1,\omega_2)$ can be evaluated as:

\vspace{-5mm}
\begin{multline}
\mathcal{L}_{I_d,I_r} (\omega_1,\omega_2)=\exp{\left\{-\lambda_s p_r t(\omega_1,\omega_2,r,d)\right\}} \\
\times \exp{\left\{-\lambda_s(1-p_r) \left[C (\omega_1^{2/\alpha}+\omega_2^{2/\alpha})+f(\omega_1,\omega_2)\right]\right\}}, \hspace{-5pt}
\label{eq:Laplace_hard1}
\end{multline}
where:
\begin{gather*}
f(\omega_1 , \omega_2) = \int_{\R^2} \frac{\omega_1 \omega_2}{(\omega_1+ ||x-d||^\alpha) (\omega_2+ ||x-r||^\alpha)} dx,\label{eq:integral_hard} \\
t(\omega_1,\omega_2,r,d)=\int_{\mathbb{R}^2}\Ex_k\left[1-z\left(\omega_1,x,k,d\right)z\left(\omega_2,x,k,r\right)\right]dx,
\label{eq:Laplace_hard2} 
\end{gather*}
and $C$ comes from (\ref{eq:C}). The expectation is with respect to the distribution $k$ of the relay and $z(\omega,x,k,d)$ is given by:
\begin{equation}
z(\omega,x,k,d)=\frac{1}{1+\omega\|x-d\|^{-\alpha}+\omega\|k-d\|^{-\alpha}}.
\label{eq:s_function}
\end{equation}
For $z(\omega,x,k,r)$ a similar expression holds interchanging $d$ with $r$. The complexity of these expressions is due mainly to the interferences (\ref{eq:power_inter_relay}) and (\ref{eq:power_inter_dest}), and it precludes closed-form computations. 
For this reason we introduce the following far-field approximation for the path loss of the interfering clusters: {the users within a cluster see the interference from other clusters as a point source of interference, meaning that:
\begin{align}
\|x_i-r\| &\approx \|x_i+k_i-r\| \approx \|x_i+\tau k_i-r\|, \\
\|x_i-d\| &\approx \|x_i+k_i-d\| \approx \|x_i+\tau k_i-d\|.
\end{align}
The parameter $\tau$ allows to establish the far field approximation using any point between each source and its relay. As we shall see the results obtained are the same independently of its value. With this assumption a single path loss will appear in the interferences, so (\ref{eq:power_inter_relay}) and (\ref{eq:power_inter_dest}) can be simplified as:}
 \begin{equation}
\tilde{I}_r= \hspace{-1mm} \sum_{i:x_i\in\Phi_s} \hspace{-1mm}\frac{|h_{x_ir}|^2+\varepsilon_{x_i}\left(|h_{k_ir}|^2+2\Re\left\{h_{x_ir}h_{k_ir}^*\rho\right\}\right)}{\|x_i+\tau k_i-r\|^{\alpha}},
\label{eq:power_inter_relay_approx}
\end{equation}
 \begin{equation}
\tilde{I}_d= \hspace{-1mm} \sum_{i:x_i\in\Phi_s} \hspace{-1mm}\frac{|h_{x_id}|^2+\varepsilon_{x_i}\left(|h_{k_id}|^2+2\Re\left\{h_{x_id}h_{k_id}^*\rho\right\}\right)}{\|x_i+\tau k_i-d\|^{\alpha}}.
\label{eq:power_inter_dest_approx}
\end{equation}
This approximation will be very good in the high reliability regime because the independent fading coefficients are conserved and the large scale effect of path loss is still taken into account.

With these new interference expressions, we upper bound the OP of the network by introducing the union bound on the outage events of DF:
\begin{multline} \label{eq:uboundhr1}
\hspace{-5pt}\poutm(R) \leq \prob\{\varepsilon_0=0\} \prob\left\{\adt(R) | \varepsilon_0 = 0\right\}  + \prob\{\varepsilon_0=1\} \\ \hspace{-7pt} \times \hspace{-1pt} \Ex_r\left[\prob\left\{\adf(R)|r,\varepsilon_0=1\right\}+\prob\left\{\bdf(R)|r,\varepsilon_0=1\right\} |\varepsilon_0 = 1\right] \hspace{-1pt}.	 \hspace{-5pt}
\end{multline}
This will be a good approximation when the relay is not too far away from the source, since in that case the event $\bdf(R)$ will be dominant and  $\adf(R)$ will have a relatively small probability of occurrence.  Using the simplified interferences we can evaluate this upper bound in closed form:

\begin{Theorem} \label{teo:opubound} The OP upper bound (\ref{eq:uboundhr1}) for this network can be evaluated as:
\begin{multline}
 \poutm(R) \leq   (1-p_r)\left[ 1- e^{- \lambda_s \Delta(p_r) D^2}\right] \\
 + p_r  \left\{ 2 - \Ex_r\left[e^{-\lambda_s \Delta(p_r) ||r||^2} \right] \right.-e^{-\lambda_s\Delta(p_r) D^2} \\ \left. 
  \!\! \left[1+  \lambda_s\Delta(p_r) D^2 \!\left( 1 + \frac{2-\alpha}{\alpha D} \Ex_r\! \left[||r-d||\right] \right) \right. \right.\\ \left. \left. 
  \phantom{\frac{1}{1}}+ \Onot \left( (\lambda_s\Delta(p_r) D^2)^2 \right) \right]	\right\},
\label{eq:poutrelrand3}
\end{multline}
as $\lambda_s \Delta(p_r) D^2 \equiv \lambda_s \delta \left(1+\frac{2p_r}{\alpha}\right) D^2 \rightarrow 0$, with:
\begin{equation}
 \Ex_r\left[e^{-\lambda_s \Delta(p_r) ||r||^2} \right] = \frac{\phi_0 \lambda_{in}}{\phi_0 \lambda_{in}+2 \lambda_s \Delta(p_r)}.
\end{equation}
When $\phi_0 = 2\pi$ we have a close form expression for the expectation:
\begin{equation}
\Ex_r\left[||r-d||\right] = \sigma_{in} Q_{2,0}(D/\sigma_{in},0), \label{eq:bouNQ20}
\end{equation}
where $Q_{2,0}$ is the $(2,0)$ Nuttall $Q$-function \cite{Nuttall1972}:
\begin{equation}
Q_{2,0} \left(s,0\right) = \sqrt{\frac{\pi}{8}} e^{-\frac{s^2}{4}} \left(\left(s^2 +2\right) I_0\left(\frac{s^2}{4}\right)+ s^2 I_1\left(\frac{s^2}{4}\right)\right) ,\nonumber
\end{equation}
where $I_0$ and $I_1$ are the modified Bessel functions of the first kind of orders $0$ and $1$. In the general case we have to find the expectation numerically or we may use the following upper bound:
\begin{equation}	
\Ex_r \left[||r-d||\right] \leq D  (1+s \gamma(s,\phi_0)) \label{eq:ineqErd}
\end{equation}
with $s = (\lambda_{in}\phi_0 D^2)^{-1/2}$ and
\begin{equation*}
\gamma(s,\phi_0) \hspace{-0.5mm}=\hspace{-0.5mm} \sqrt{\frac{\pi}{2}} \left\{1+ \left[\frac{8(1-\cos(\phi_0/4))}{\phi_0}-2\right] \erf\left(\frac{1}{\sqrt{2}s}\right)\right\}.
\end{equation*}
\end{Theorem}
\begin{IEEEproof}
See appendix \ref{proof:opubound}.
\end{IEEEproof}

\section{Optimal relay activation probability}
\label{subsec:randomrelays}
In the previous section we established an upper bound on the OP of the network choosing the relay as the NN in a cone, as a function of the relay activation probability $p_r$ and the cone aperture $\phi_0$. For a given network set-up ($R$, $\alpha$, $\lambda_s$, $\sigma_{in}$) different values of $p_r$ and $\phi_0$ will yield different values of the OP:  increasing $p_r$ will introduce additional interference in the network, while decreasing the cone aperture $\phi_0$ will increase the average source-relay distance. If there is a high density of potential relays the cone aperture can be used to balance the average source-relay and source-destination distances to optimize the performance of the network. For this reason we should find the optimal values of $p_r$ and $\phi_0$, those which result in the smallest OP for each setup. In this section we study the optimal value of $p_r$ in terms of the OP and determine the gains that can be achieved in terms of OP by optimizing this parameter.

Optimization of the relay activation probability using standard methods is very involved due to the non-linear nature of the expression of the OP. It would be expected that an optimal relay activation probability $p_r$ would exist, which would optimally balance the effect of the added interference and the gains of activating additional relays.
\begin{Theorem}[Concavity of the OP] \label{lemma:concavOPrr}
Neglecting the term $\Onot \left( (\lambda_s \Delta(p_r) D^2)^2 \right)$ in (\ref{eq:poutrelrand3}), for each network set-up ($\alpha$, $d$, $\phi_0$, $\lambda_s$, $R$) such that $\lambda_s \delta D^2 < 0.38$ there is an interval of $\sigma_{in}$:
\begin{equation}
0 \leq \sigma_{in} \leq \sigma_c,
\end{equation}
such that the OP upper bound is a concave function of $p_r$. 
\end{Theorem}
\begin{IEEEproof}
	See appendix \ref{proof:concavOPrr}.
\end{IEEEproof}

\begin{Lemma} \label{lem:rocke}
Given a concave function $h(x)$ in a bounded and closed interval $[x_1,x_2]$, its minimum is attained at $x_1$ or $x_2$.
\end{Lemma}
\begin{IEEEproof}
See Theorem 32.1 in \cite{rockafellar_convex_1970}.
\end{IEEEproof}
Using lemma \ref{lem:rocke} together with theorem \ref{lemma:concavOPrr} we conclude that the best OP performance for any cluster in the network can be attained when all  ($p_r = 1$) or none  ($p_r = 0$) of the sources decide to use their associated relays. In one case all the clusters will be using DF and in the other case all of them will be using DT.  This is a somewhat surprising result in the sense that in terms of the OP the best performance can be obtained either by full cooperation or by not cooperating at all. There is no ``optimal'' density of used relays in the network or optimal mixed behavior in the sense that some clusters would enjoy the advantages of cooperation while others use DT, in order to balance the generated interference. This interval clearly depends on the network set-up parameters; however, working in the realistic high reliability regime we can obtain a simple approximation of this condition that depends only on basic network parameters:
\begin{corollary} \label{coro:concavcond}
In the high reliability regime, we approximate the concavity interval of theorem \ref{lemma:concavOPrr} by finding the smallest positive solution to the equation:
\begin{equation}
\frac{4\pi \alpha \sigma_{c}^2}{\phi_0  D^2} +\left(\alpha-2\right) \frac{\Ex_r[||r-d||]}{D} - \alpha = 0.
\label{eq:concavcond}
\end{equation}
Notice that the expectation also depends on $\sigma_{c}$ so the equation cannot be solved in closed form. By using (\ref{eq:ineqErd}) to upper bound the expectation, the following sufficient condition for a concave OP concave is obtained:
\begin{equation}
\sigma_{in} \leq D \sqrt{\frac{\phi_0}{2 \pi}}\left\{ \left[\frac{1}{\alpha}+\varphi_c(\phi_0,\alpha)^2\right]^{1/2}
\hspace{-8pt}-\varphi_c(\phi_0,\alpha)\right\}, \label{eq:condconvclosed}
\end{equation}
where:
\begin{equation}
\varphi_c(\phi_0,\alpha) = \frac{1}{4} \left(1-\frac{2}{\alpha}\right) \gamma(1/\sqrt{2},\phi_0).
\end{equation}
\end{corollary}
\begin{IEEEproof}
See appendix \ref{proof:concavOPrr}.
\end{IEEEproof}
So far we have established that there is a regime in which either $p_r = 1$ or $p_r = 0$ are the values that minimize the OP for a given network set-up ($\alpha$, $d$, $\phi_0$, $\lambda_s$, $R$), and in corollary \ref{coro:concavcond} we have determined conditions to find that interval.
Now we wish to establish conditions under which we should activate all the relays, that is, when $p_r = 1$ will be the optimal choice:	
\begin{Theorem}[Optimality region of $p_r=1$] \label{lemma:optOPrr}
Neglecting the term $\Onot \left( 	(\lambda_s \Delta(p_r) D^2)^2 \right)$ in (\ref{eq:poutrelrand3}), for each network set-up ($\alpha$, $d$, $\phi_0$, $\lambda_s$, $R$) such that $\lambda_s \delta D^2 < 0.38$ there is an interval of $\sigma_{in}$:
\begin{equation}
0 \leq \sigma_{in} \leq \sigma_t
\end{equation}
such that the OP upper bound is minimized by activating all the relays.

For the high reliability regime, an approximation for $\sigma_t$ is obtained by finding the smallest positive solution of the equation:
\begin{equation}
 \left[1+\frac{2}{\alpha}\right] \left[\frac{4 \pi \sigma_t^{2}}{\phi_0 D^2}  + \left(1-\frac{2}{\alpha}\right)\frac{\Ex_r[||r-d||]}{D}\right] =1. \label{eq:condpr1opt}
\end{equation}
Notice that the expectation also depends on $\sigma_t$ so the equation is coupled. By using (\ref{eq:ineqErd}) to upper bound the expectation, the following sufficient condition for $p_r = 1$ to be optimal is obtained:
\begin{equation} \label{eq:pr1opt2}
\hspace{-2pt}\sigma_{in} \leq D \sqrt{\frac{\phi_0}{2 \pi}}\left\{ \hspace{-3pt} \left[\frac{2}{\alpha(\alpha+2)}+\varphi_t(\phi_0,\alpha)^2\right]^{1/2}
\hspace{-13pt}-\varphi_t(\phi_0,\alpha)\right\} \hspace{-3pt}, \hspace{-9pt}
\end{equation}
where:
\begin{equation}
\varphi_t(\phi_0,\alpha) = \frac{1}{4} \left(1-\frac{2}{\alpha}\right) \gamma(1/2,\phi_0).
\end{equation}
\end{Theorem}
\begin{IEEEproof}
See appendix \ref{proof:optOPrr}.
\end{IEEEproof}

Using the previous theorems, we are able to state a relay activation scheme that optimizes the OP in a network operating in the high reliability regime: for a given value of $\phi_0$ if $\sigma_{in}$ is less than the solution of (\ref{eq:condpr1opt}) then all the relays should be on. Otherwise, the relays should be turned off and DT should be employed. A computationally simpler alternative for turning the relays on would be using condition (\ref{eq:pr1opt2}) instead.
The value of $\phi_0$ could additionally be chosen within this scheme to minimize the OP. Notice that $\sigma_t \equiv \sigma_t(\phi_0)$ is a function of $\phi_0$. If for a network set-up ($\alpha$, $d$,  $\lambda_s$, $R$) there is a value of $\phi_0$ such that $p_r = 1$ is optimal, i.e. $\sigma_{in} \leq \sigma_t(\phi_0)$ holds, then there will be a range of values of $\phi_0$ for which this condition will hold. We should therefore choose the value of $\phi_0$ for which $\sigma_{in} < \sigma_t(\phi_0)$ holds and the OP is minimized. On the other hand, if there is no value of $\phi_0$ such that $\sigma_{in} < \sigma_t$ we have that $p_r =0 $ is optimal and hence, DT should be employed.

As we shall observe in the section of numerical results, there will be scenarios in which setting $\phi_0 = 2 \pi$ will yield approximately the same performance as optimizing the value of $\phi_0$ in terms of the OP according to the previous observation. This means that in practical scenarios, this optimization may not always be of importance and the value of $\sigma_t$ can be obtained by setting $\phi_0 = 2 \pi$ in (\ref{eq:condpr1opt}).

\begin{figure}[t!]
\centering
	\ifpdf
		\includegraphics[width=.95\columnwidth,keepaspectratio,trim= 0mm 1mm 10mm 5mm,clip]{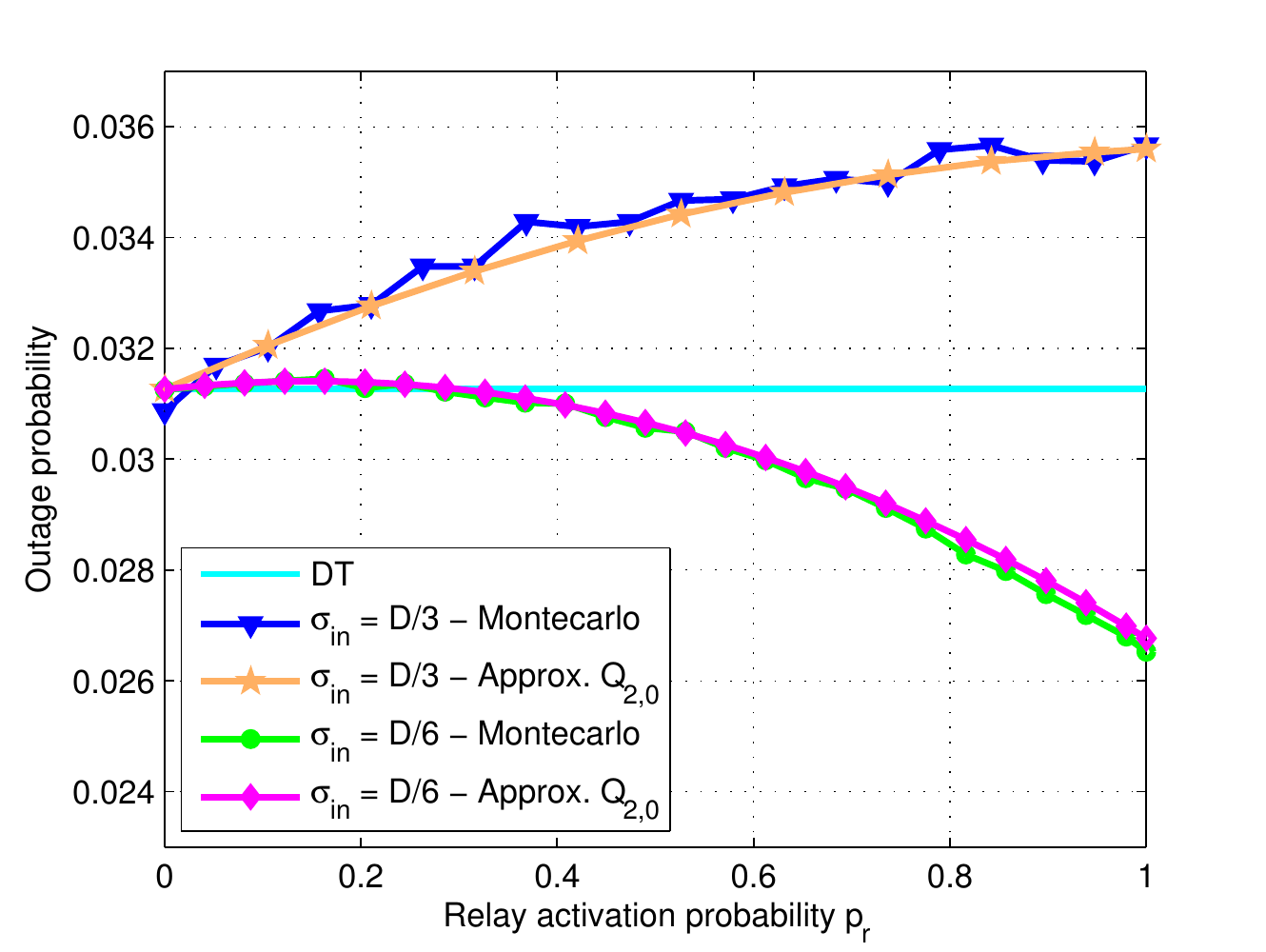} 
	\else
		\includegraphics[width=.95\columnwidth,keepaspectratio,trim= 0mm 1mm 10mm 5mm,clip]{poutrandrel-def2.eps} 
	\fi
	\caption{Outage probability $\poutm(R,p_r)$ as a function of $p_r$ for values of $\sigma_{in}$ showing optimality of $p_r=0$ or $p_r = 1$. $d=(10,\ 0)$, $\lambda_s=10^{-4}$, $R=0.5$, $\alpha = 4$. Montecarlo simulations are obtained by averaging $8 \times 10^6$ realizations of the PPP using (\ref{eq:power_inter_relay}) and (\ref{eq:power_inter_dest}). Approximations come from using (\ref{eq:bouNQ20}) in (\ref{eq:poutrelrand3}).}
	\label{fig:concavop}
\end{figure}

Finally, we want to compare the OP that can be achieved with the scheme defined in theorem \ref{lemma:optOPrr}  with the one obtained using only DT. For each value of $\phi_0$, if $p_r = 1$ minimizes the OP then the scheme will exhibit gains with respect to DT, while if $p_r = 0$ is the optimum, the scheme reverts to DT, and no gains will be seen. The following theorem finds the approximate reduction of the OP of the scheme with respect to DT:
\begin{Theorem} \label{teo:opgain}
In the high reliability regime, the relative decrease in OP of the activation scheme obtained by using (\ref{eq:condpr1opt}) is:
\begin{equation} \label{eq:mixvsdt}
\hspace{-5pt} \frac{\poutm}{\poutdt} \hspace{-1pt} \approx\hspace{-1pt}  \begin{cases}
\left(1+\frac{2}{\alpha}\right) \hspace{-2pt} \left(\frac{4\pi \sigma^2_{in}}{\phi_0 D^2} \hspace{-2pt}+ \hspace{-2pt} \left(1-\frac{2}{\alpha}\right)  \frac{\Ex_r\left[||r-d||\right]}{D}\right) &\hspace{-5pt} \text{\small$\sigma_{in} \leq \sigma_t$,} \\
1 & \hspace{-5pt}\text{\small otherwise.} \\
\end{cases}
\end{equation}
In addition $\sigma_t$ can be lower bounded by (\ref{eq:pr1opt2}).
\end{Theorem}
\begin{IEEEproof}
See appendix \ref{proof:opgain}.
\end{IEEEproof}
As we mentioned before, both $\sigma_t$ and the actual reduction in OP are a function of $\phi_0$. If $\phi_0$ can be optimized, then for each network setup ($\alpha$, $d$, $\lambda_s$, $R$) we have to determine (if they exist) the values of $\phi_0$ such that $\sigma_{in} \leq \sigma_t$ (which ensure a gain with respect to DT) and from those values, the one that  minimizes the OP.

Finally, it is interesting to mention that under certain conditions the OP is not a concave function of $p_r$, that is, the OP is minimized by choosing a value of $p_r$ which is different from $p_r = 0$ or $p_r = 1$. However, in such scenarios the network is well outside the high reliability regime and typical operating conditions.

\section{Numerical results} \label{sec:numerical}
In this section we present some simulations to study the behavior of the expressions we have introduced previously. In Fig. \ref{fig:concavop} the OP with respect to $p_r$ is plotted for two different values of $\sigma_{in}$, one in which $p_r=1$ is optimal and another one for which $p_r =0$ is the optimal point, when the relay is selected as the nearest neighbor on the whole plane ($\phi_0 = 2\pi$). The theoretical expressions come from the upper bound (\ref{eq:poutrelrand3}) using (\ref{eq:bouNQ20}), and they are compared with Montecarlo simulations obtained by averaging $8 \times 10^6$ realizations of the PPP using the true interferences (\ref{eq:power_inter_relay}) and (\ref{eq:power_inter_dest}), taking $d=(10,\ 0)$, $\lambda_s=10^{-4}$, $R=0.5\text{ bit/use}$ and $\alpha =4$. We see that the approximations derived with the simplified interferences (\ref{eq:power_inter_relay_approx}) and (\ref{eq:power_inter_dest_approx}) are in excellent agreement with the actual OP derived with the more complex interferences.

\begin{figure}[!t]
\centering
	\ifpdf
		\includegraphics[width=.95\columnwidth,keepaspectratio,trim= 0mm 3mm 10mm 5mm,clip]{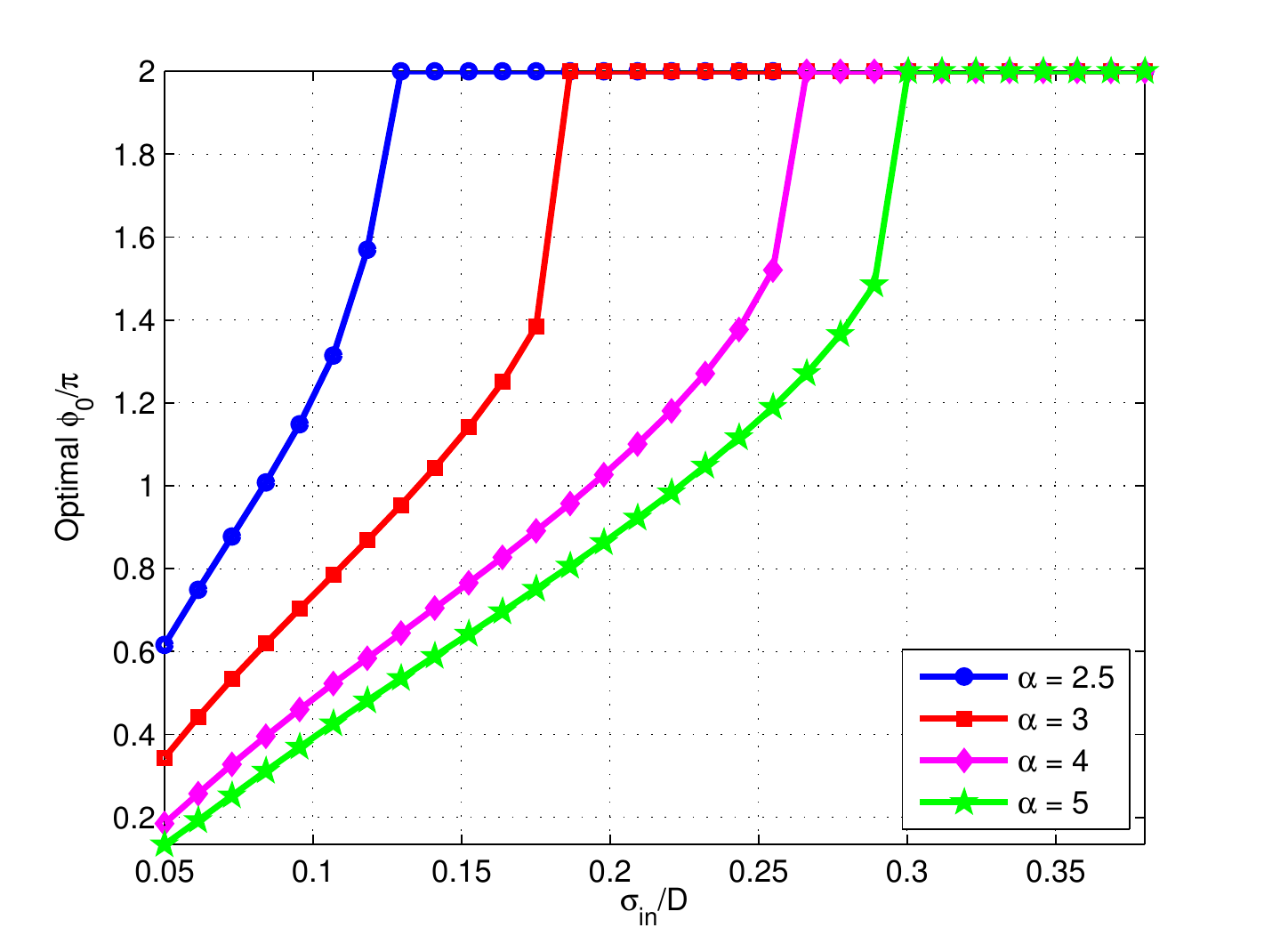} 
	\else
		\includegraphics[width=.95\columnwidth,keepaspectratio,trim= 0mm 3mm 10mm 5mm,clip]{Phi0opt3.eps} 
	\fi
\caption{Optimal cone aperture $\phi_0$ as a function of $\sigma_{in}/D$ obtained using (\ref{eq:mixvsdt}) for different values of $\alpha$. $\lambda_s = 10^{-4}$. $d = (10 , 0)$.}
\label{fig:optphi0}
\end{figure}

In Fig. \ref{fig:optphi0} we plot the optimal cone aperture $\phi_0$ as a function of $\sigma_{in}/D$ for different values of the path loss exponent $\alpha$ and for $d = (10, 0)$. 
To do this, we numerically  find the value of $\phi_0$ that maximizes the OP gain of the mixed protocol with respect to DT  in (\ref{eq:mixvsdt}) for each value of $\sigma_{in}/D$. It is interesting to note that as the path loss exponent decreases the optimal cone aperture becomes $\phi_0 = 2\pi$ for a large range of values of $\sigma_{in}/D$. Only when the network of potential relays is very dense (small $\sigma_{in}/D$) a value of $\phi_0 < 2\pi$ should be chosen. This is because when the exponent diminishes both the source-relay and the interference paths become stronger, but the effect of the increased interference is dominant. Thus the diminished exponent creates an effect equivalent to increasing the average source-relay distance. The value of $\phi_0$ must therefore become larger in order to decrease the average source-relay distance and compensate for this effect. 

\begin{figure}[t]
\centering
	\ifpdf
		\includegraphics[width=.95\columnwidth,keepaspectratio,trim= 0mm 0mm 10mm 5mm,clip]{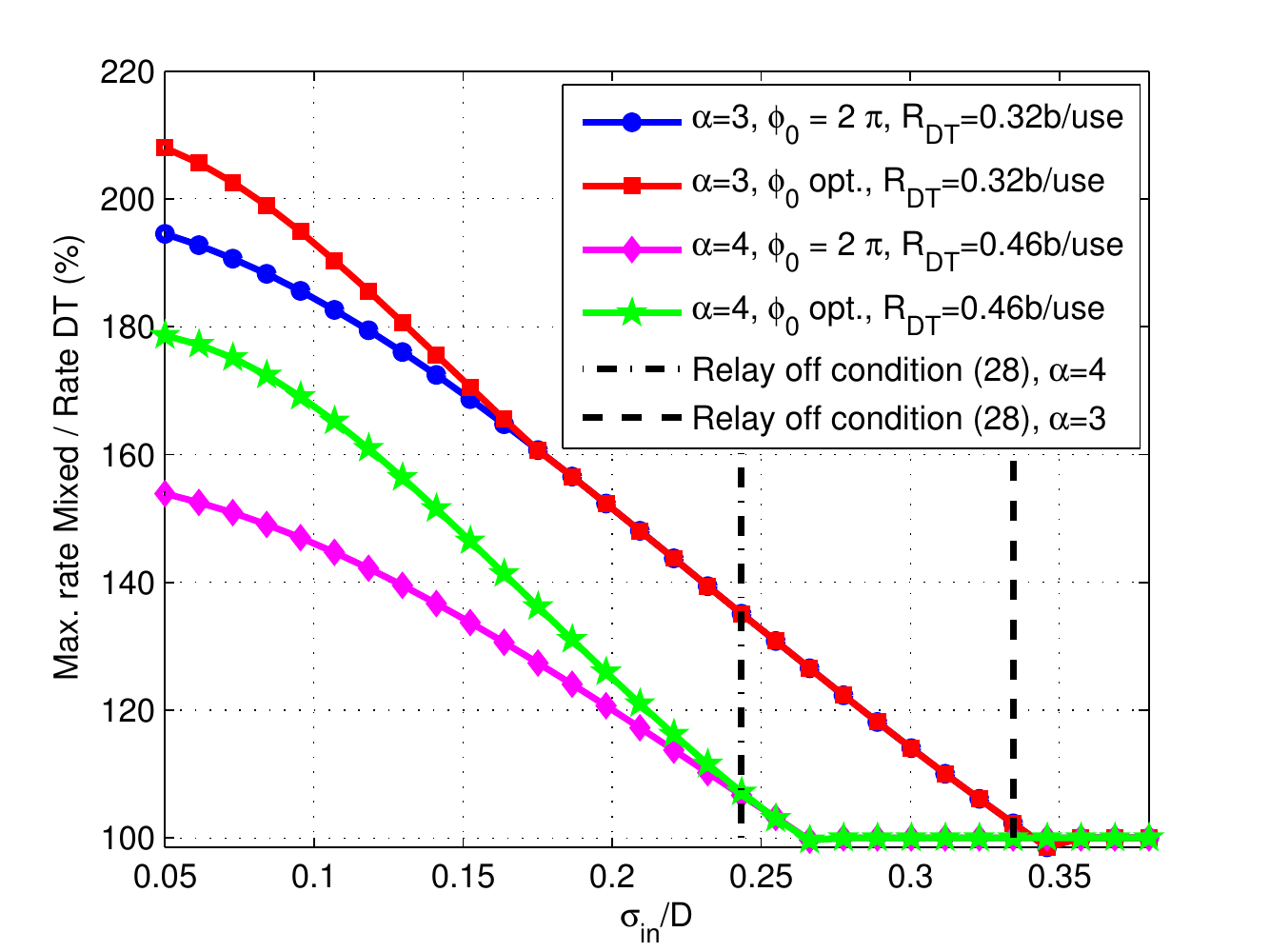} 
	\else
		\includegraphics[width=.95\columnwidth,keepaspectratio,trim= 0mm 0mm 10mm 5mm,clip]{Maxrate2.eps} 
	\fi
	\caption{Maximum rate attainable for the on/off relaying strategy relative to the same rate of DT for a given OP constraint. The mixed scheme rates are obtained by using (\ref{eq:poutrelrand3}) and the on/off condition from solving (\ref{eq:condpr1opt}). We also plot the on/off condition (\ref{eq:pr1opt2}). The optimal aperture angles come from Fig. \ref{fig:optphi0}. $\lambda_s = 10^{-4}$. $d = (10 , 0)$.}
	\label{fig:maxrate}
	\end{figure}

In Fig. \ref{fig:maxrate} we study the maximum rate attainable for the on/off relaying strategy relative to the same rate of DT  in percentage for a desired OP value of $0.03$. The maximum rates are obtained by using (\ref{eq:poutrelrand3}).
For the plots with $\phi_0= 2\pi$ the rates are obtained by using (\ref{eq:bouNQ20}) while in the other case the expectations are computed numerically. For the plots with optimized cone aperture we use the values of $\phi_0$ from Fig. \ref{fig:optphi0}, taking $\lambda_s = 10^{-4}$ and $d = (10 , 0)$. The on/off condition  (which predicts when the rate of the mixed scheme reaches that of DT) is obtained by solving (\ref{eq:condpr1opt}). We have also plotted as vertical lines the simpler on/off condition (\ref{eq:pr1opt2}) which is in excellent agreement with the other one. We observe that optimizing the cone aperture can be helpful when the path loss exponent or the density of potential relays are large. In addition, as the path loss exponent decreases we can achieve a lower maximum rate with DT for a given outage constraint; this implies that the benefits of a reduced exponent within the cluster are outmatched by the simultaneous increase in interference due also to the reduced exponent. The plot also shows that although the maximum rate for DT is smaller, the relative gains of the mixed scheme become larger. This means that the maximum achievable rate decreases slower for the mixed scheme than for DT as the path loss exponent decreases, which suggests that the increased interference is less damaging for the mixed scheme than for DT.

\begin{figure}[!t]
		\ifpdf
			\includegraphics[width=.95\columnwidth,keepaspectratio,trim= 0mm 1mm 10mm 5mm,clip]{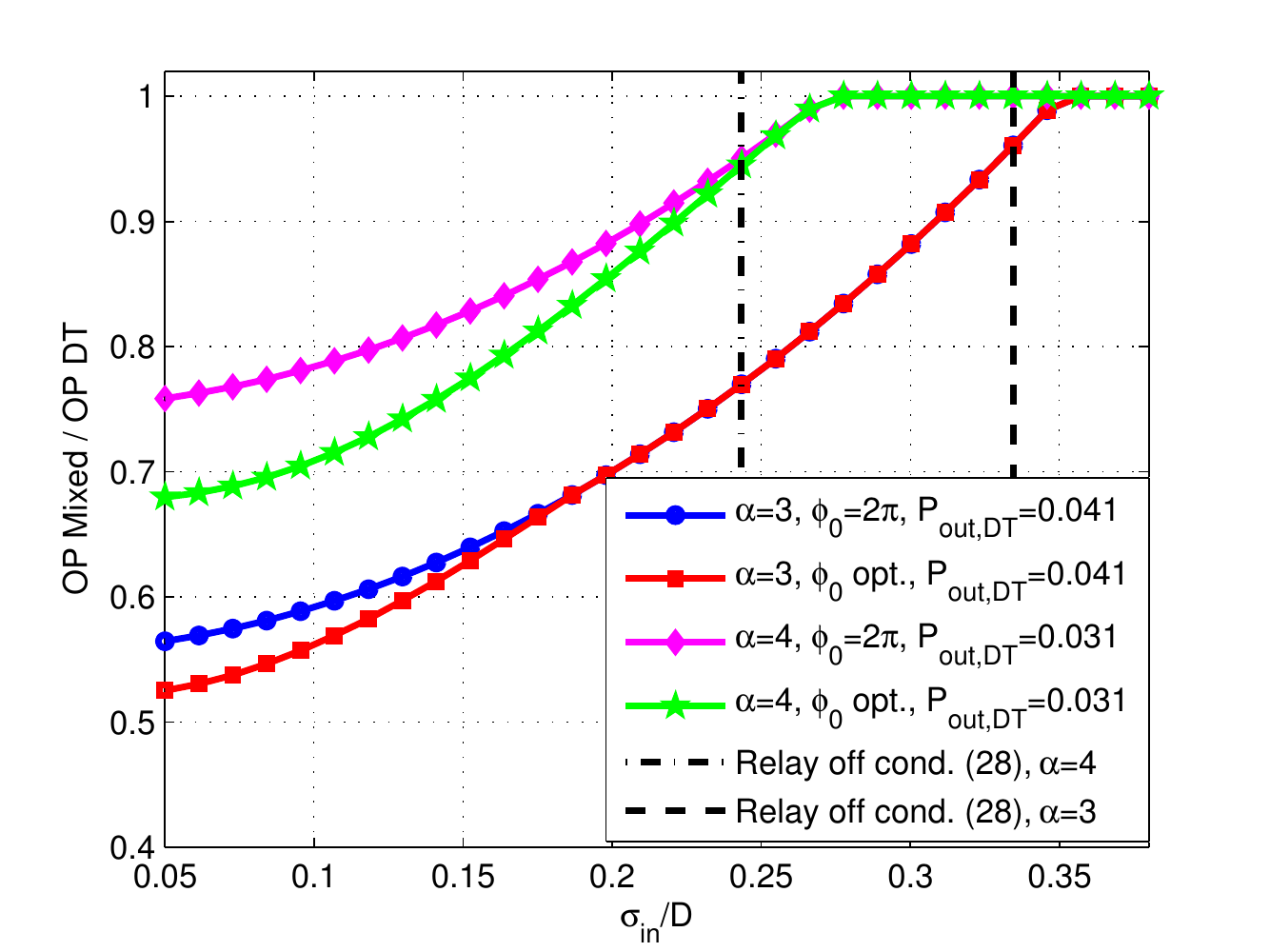} 
		\else
			\includegraphics[width=.95\columnwidth,keepaspectratio,trim= 0mm 1mm 10mm 5mm,clip]{OPopt2.eps} 
		\fi
		\caption{Relative improvement in OP with respect to DT for the on/off scheme as a function of $\sigma_{in}/D$ as predicted by theorem \ref{teo:opgain}. We have also plotted as vertical lines the on/off condition (\ref{eq:pr1opt2}). $d = (10, 0)$.}
		\label{fig:Opgains}
\end{figure}

In Fig. \ref{fig:Opgains}  we plot the relative gain in OP with respect to DT as a function of $\sigma_{in}/D$ using $\phi_0 = 2\pi$ and the optimal cone apertures from Fig. \ref{fig:optphi0}. The OP gains are obtained from theorem \ref{teo:opgain}. We also plotted as vertical lines the simpler condition (\ref{eq:pr1opt2}) which is in excellent agreement with the other one.


\begin{figure}[!t]
\centering
	\ifpdf
		\includegraphics[width=.95\columnwidth,keepaspectratio,trim= 0mm 0mm 0mm 0mm,clip]{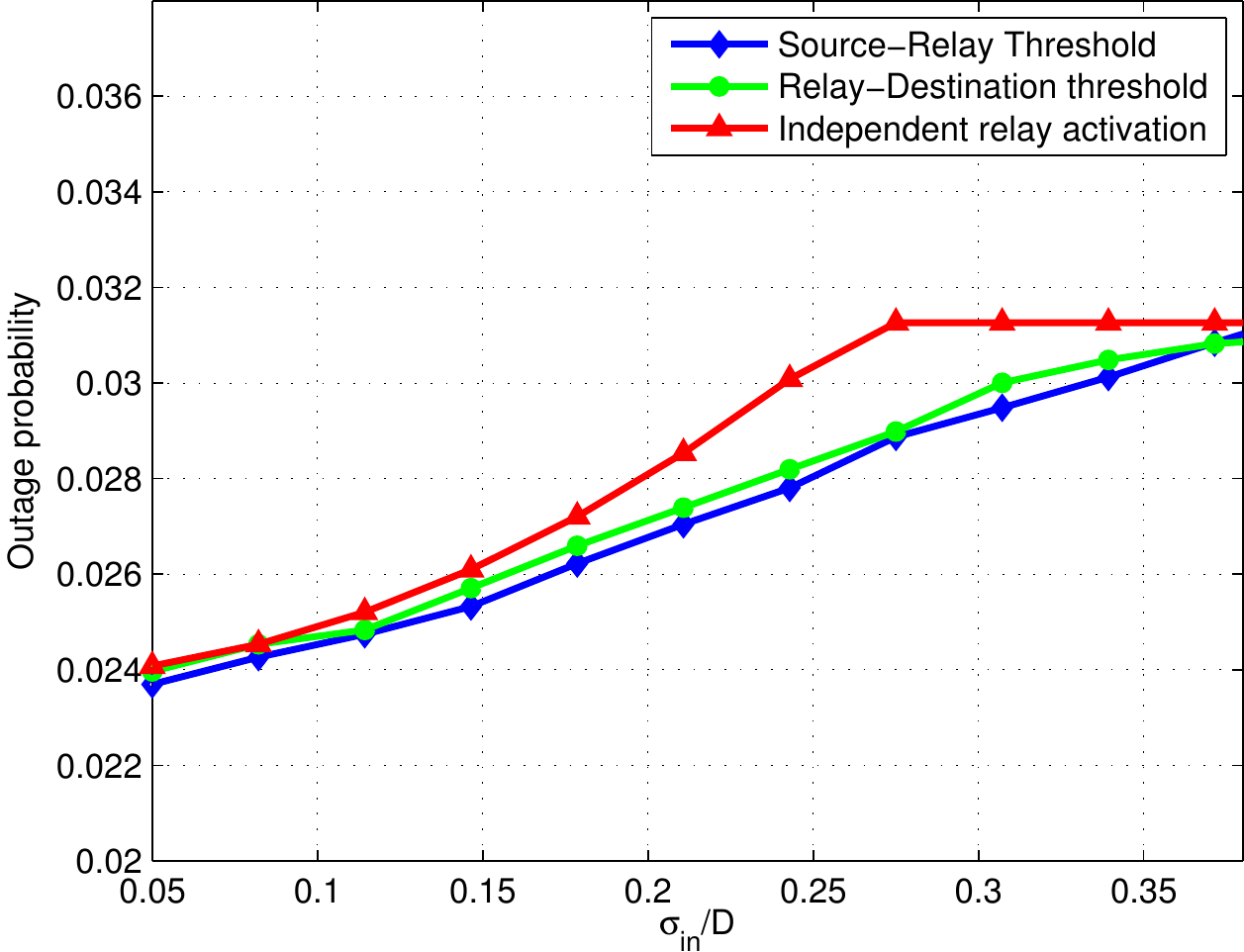} 
	\else
		\includegraphics[width=.95\columnwidth,keepaspectratio,trim= 0mm 0mm 0mm 0mm,clip]{Poutthall.eps} 
	\fi
\caption{Comparison between the OP attainable through independent relay activation and through the use of a threshold on the source-relay or source-destination channel. $\lambda_s = 10^{-4}$, $R=0.5$ b/use, $d = (10 , 0)$, $\alpha = 4$. The independent relay curve comes from (\ref{eq:poutrelrand3}) and the rest from Montecarlo simulations of the PPP. Thresholds are optimized numerically for best performance.}
\label{fig:stratcomp}
\end{figure}

Finally, in Fig. \ref{fig:stratcomp} we compare the performance of the proposed on/off  strategy against with two other simple relay activation schemes: one in which the relay is activated if the source-relay channel exceeds a threshold and another one in which a threshold on the relay-destination channel is used. Both schemes make use of the available CSI. In the first case, the relay can determine if the threshold is exceeded, and in the second one, the destination, who has CSI on the relay-destination link, can send a bit (at negligible cost) indicating if the relay should transmit or not. In both cases, the path loss and the corresponding fading coefficients are considered. The OP curves for these schemes are determined through Montecarlo simulations of the point process and for each point the value of the threshold is numerically optimized to obtain the smallest OP possible. These curves are compared to the OP from the upper bound (\ref{eq:poutrelrand3}) and the on/off strategy. For these simulations we use $\lambda_s = 10^{-4}$, $R=0.5$ b/use, $d = (10 , 0)$, $\alpha = 4$. We observe that although these schemes employ available CSI which is not taken into account by the independent activation schemes, the performance is similar between the three strategies.

\section{Summary and Final Remarks}
\label{sec:final}

In this work we analyzed the performance of a  large wireless network under a mixed cooperative randomized scheme which employs either DF or DT, and obtained the optimal relay activation strategy for this network. When DF is used, the relays are chosen as the nearest neighbor within a cone, with its axis towards the destination. This is a natural assumption since DF is known to be near optimal when the relay is not too far from the source. At the same time, the effect of the path loss on the relay-destination link, which is very detrimental to the performance of the scheme, is reduced. The choice between DT and DF is done by the corresponding relay associated with each source via a randomized decision with probability $p_r$ and without taking into account any additional knowledge the relays might have. This simple procedure, which is mathematically tractable, can be thought as a MAC layer at the relays (in a similar fashion as the popular ALOHA protocol), with the objective of limiting the interference generation in the network. On the other hand it could also model a situation in which the relays are unavailable due to conditions out of control of the source or the relay itself, such as, for example, a depleted battery. With this simple model, a balance between cooperation and interference generation can be established in the network. Surprisingly, for typical operating conditions, the optimal values of $p_r$ are 0 or 1, revealing a binary behavior:  all nodes in the network should use their relay or none at all. Following this conclusion, a relay activation strategy was introduced to achieve the optimal behavior. Even when cooperation is beneficial to all, the performance improvements may not be as large as in the typical fading relay channel with Gaussian noise. The reason for this comes from the fact that, in addition to fading, we have averaged over all possible node configurations, including many cases in which interference is very damaging. It is interesting to mention that the model introduced and several results, such as theorem \ref{teo:opubound}, can be used to study other relay selection and activation algorithms based on position, such as choosing the relay as the nearest or farthest neighbor on a finite cone, and with minor modifications extend them to other cases involving additional CSI. Other protocols assuming higher degrees of CSI may yield better gains, but this may not be a realistic assumption in this context. 
A potential improvement could be obtained using more sophisticated cooperative transmission schemes which could take into account the impairments generated by the nearby interferers \cite{Baccelli-ElGamalTse11} (which introduce by far the most harmful interference). Basically, this could consist on decoding the messages sent by strong nearby interferers first, then subtracting them from the received signal and finally attempting to decode the desired message. In such situation, besides the intrinsic benefits of cooperation, the smart use of the aggregate interference introduced in part by the cooperating nodes, could ameliorate its harmful effect on the overall network.  Another improvement could be obtained by using more advanced MAC schemes for the relays and the sources, such as CSMA, to avoid simultaneous nearby transmissions. In this context a metric such as the transmission capacity \cite{weber_overview_2010} may be more appropriate for the analysis.
Finally, the study of other cooperative schemes as AF and CF deserves full consideration. All these issues, as well as the effect of using several potential relays instead of only one, constitute important and interesting future work directions.

\appendix
\section{}
\subsection{Interference RVs and their Laplace transforms} \label{ap:Laptrans}
This appendix is a simple review of the basic properties of the LT used in this work. For details see \cite{BB2010} and \cite{DVJ1998}. Let $\tilde{\Phi} = \left\{\left(x_i,m\right)\right\}$ be an independently marked homogeneous PPP with $\Phi = \left\{x_i\right\}$ the homogeneous PPP in $\R^2$ and $m$ a vector of marks on a subset of $\R^l$, $l\geq 1$. Define the interference RVs:
\begin{alignat}{2}
I_d = \sum_{i:x_i\in\Phi_s} f_1(d,x_i,m_i),\hspace{2mm} & \hspace{2mm} I_r = \sum_{i:x_i\in\Phi_s} f_2(r,x_i,m_i),
\end{alignat}
where $f_1$ and $f_2$ are real valued non negative functions. The joint LT of the interference RVs at $(\omega_1,\omega_2)$ is  \cite{kingman93} \cite{BB2010}:
\begin{multline}
\hspace{-5pt} \mathcal{L}_{I_d,I_r}(\omega_1,\omega_2) =\\ 
\hspace{-3pt} \exp \hspace{-2pt} \left\{ -\lambda_s \hspace{-3pt} \int_{\R^2} \hspace{-3pt} \Ex_m\left[1- e^{ -\omega_1f_1(d,x,m) - \omega_2f_2(r,x,m)}\right]dx\right\} \hspace{-2pt}.\hspace{-3pt}  \label{eq:Lapgen1}
\end{multline}
Taking $\omega_1=0$ or $\omega_2=0$ the single LT are obtained.
\begin{Lemma} \label{lemma:Lapcase2}
Suppose the marks of the HPPP are $m = \left(|h_1|^2,|h_2|^2,\varepsilon,k\right)$, with $|h_1|^2$ and $|h_2|^2$ unit mean independent exponential RVs, $\varepsilon$ a Bernoulli RV with success probability $p_r$, and $k$ a RV on $\R^2$. Let $f_1(d,x,m) = \left(|h_1|^2 + \varepsilon |h_2|^2\right) l(x+\tau k,d)$ with the path loss function $l(x,y) = ||x-y||^{-\alpha}$ and $\tau \in [0,1]$. Then the LT is:
\begin{equation}
\mathcal{L}_{I_d}(\omega_1) = \exp\left\{-\lambda_s C \omega_1^{2/\alpha} \left(1+\frac{2p_r}{\alpha}\right)\right\}. \label{eq:laptrans}
\end{equation}
\end{Lemma}
\begin{IEEEproof}
Taking $\omega_2 = 0$ in (\ref{eq:Lapgen1}), writing the expectation with respect to the marks and interchanging the integration order we find that:
\begin{multline}
\hspace{-12pt}\mathcal{L}_{I_d}(\omega_1)\hspace{-2pt} = \hspace{-2pt}\exp \hspace{-2pt} \left\{\hspace{-3pt} -\lambda_s p_r \hspace{-2pt} \int_{\R^2} \hspace{-2pt} \int_{\R^2} \hspace{-3pt}  1-\frac{1}{\left[1+\omega_1 l(x+\tau k,d)\right]^2}dx dF_k\right.\\
\left. \hspace{-7pt} -\lambda_s (1-p_r)  \hspace{-2pt}\int_{\R^2} \hspace{-2pt} \int_{\R^2} \frac{1}{1+(\omega_1 l(x+\tau k,d))^{-1}}dx dF_k
\right\}\hspace{-2pt} .\hspace{-10pt} 
\end{multline}
When the integrals with respect to $x$ are computed the result does not depend on $k$ so the distribution of $k$ does not affect the final result. For the first integral we have:
\begin{align}
\hspace{-2pt} \int_{\R^2}\hspace{-3pt}  1-\frac{1}{\left[1+\omega_1 l(x+\tau k_i,d)\right]^2}dx &=2\pi\omega^{2/\alpha} \hspace{-2pt}\int_{0}^{\infty} \hspace{-4pt}\frac{1+2t^\alpha}{(1+t^\alpha)^2}tdt \nonumber\\
&=\omega_1^{2/\alpha} \left(1+\frac{2}{\alpha}\right)C.
\end{align}
For the last step we integrate by parts and $C$ is defined in (\ref{eq:C}). The second integral is known from the DT case \cite{baccelli_aloha_2006}:
\begin{equation}
\int_{\R^2} \frac{1}{1+(\omega_1 l(x,r))^{-1}}dx = C\omega_1^{2/\alpha}. \label{eq:intbacc1}
\end{equation}
\end{IEEEproof}

\vspace{-2mm}
\subsection{Proof of lemma \ref{lemma:tightness}} \label{proof:tightness}
The interference signals at the relay and the destination are:
\begin{align}
\tilde{Z}_r &= \sum_{i:x_i\in\Phi_s}\!\!\!\!\left(\!\frac{h_{x_ir}X_{x_i}}{\|x_i-r\|^{\frac{\alpha}{2}}}+\varepsilon_{x_i}\frac{h_{k_ir}X_{k_i}}{\|x_i+k_i-r\|^{\frac{\alpha}{2}}}\!\right) \\
\tilde{Z}_d &= \sum_{i:x_i\in\Phi_s}\left(\!\frac{h_{x_id}X_{x_i}}{\|x_i-d\|^{\frac{\alpha}{2}}}+\varepsilon_{x_i}\frac{h_{k_id}X_{k_i}}{\|x_i+k_i-d\|^{\frac{\alpha}{2}}}\!\right)
\end{align}
where $(X_{x_i},X_{k_i})$ are the complex, circular and zero-mean Gaussian signals with correlation coefficient $\rho$ of each source and its relay (if it's active) \cite{kramer_cooperative_2005}. The proof of the lemma follows from the fact that when $\alpha>2$, $I_d$ and $I_r$ are finite for almost every realization of $\tilde{\Phi}_s$. This can be shown using the Laplace functional of $\tilde{\Phi}_s$ \cite{BB2010}, \cite{haenggi_stochastic_2009}, and the following functions 
\begin{multline}
f(x,\varepsilon_,k,h_{1},h_{2},h_{3},h_{4})=\frac{|h_{1}|^2}{\|x-r\|^{\alpha}}
\\ +\varepsilon\left(\frac{|h_{3}|^2}{\|x+k-r\|^{\alpha}}\right.
\left.+\frac{2\Re\left\{h_{1}h_{3}^*\rho\right\}}{\|x-r\|^{\frac{\alpha}{2}}\|x+k-r\|^{\frac{\alpha}{2}}}\right)
\end{multline}
and $g(x,\varepsilon,k,h_{1},h_{2},h_{3},h_{4})$ defined in an similar form.
Since signaling between clusters is correlated with correlation coefficient $\rho$ within the cluster and independent between clusters, it can be shown that the partial sums (through a proper enumeration of the points of the particular realization of $\tilde{\Phi}_s$) in  $\tilde{Z}_r$ and $\tilde{Z}_d$ are Gaussian with variances given by the corresponding partial sums in $I_d$ and $I_r$. Thanks to the finiteness of $I_d$ and $I_r$, and the \emph{tightness} property (Theorem 25.10 in \cite{billingsley_probability_1995}) we have the desired result.

\vspace{-2mm}
\subsection{Proof of theorem \ref{teo:opexp}} \label{proof:opexp}
Define $V:= |{h}_{sd}|^2 l_{sd}+|{h}_{rd}|^2 l_{rd}$ and write:
\begin{multline}
\prob \left\{\adf(R) \cup \bdf(R)  |r, \varepsilon_0 = 1 \right\}=  \\= 1 - \prob \left\{|h_{sr}|^2 \geq \frac{TI_r}{l_{sr}},V\geq TI_d \biggl|{r, \varepsilon_0 = 1}\right\}\\
= 1 - \Ex_{\tilde{\Phi}_s} \left[\bar{F}_{h_{sr}}(TI_r/l_{sr}) \bar{F}_{V}(TI_d)\right], \label{eq:PoutDF1}
\end{multline}
where $\bar{F}_{h_{sr}}(u) = e^{-u}$ and $\bar{F}_{V}(\cdot)$ are the complementary cumulative distribution function (CCDF) of $h_{sr}$ and $V$, respectively, and we used that $h_{sr}$ and $V$ are independent of each other and of $\tilde{\Phi}_s$. Since $\rho = 0$, when $||r-d|| \neq D$, $V$ is distributed as the sum of two independent exponential RVs with different means. In that case, its CCDF is:
\begin{equation}
\bar{F}_{V}(u) = \frac{D^\alpha e^{-u ||r-d||^\alpha}-||r-d||^\alpha e^{- u D^\alpha}}{D^\alpha-||r-d||^\alpha }.
\end{equation}
When $||r-d||= D$, the means of the exponential RVs are the same so $V$ follows a Gamma distribution with $2$ degrees of freedom. However this does not affect the average with respect to $r$ that we need to carry out. Replacing both CCDFs in (\ref{eq:PoutDF1}) we obtain $\prob \left\{\adf(R) \cup \bdf(R)  |r, \varepsilon_0 = 1 \right\}$ from (\ref{eq:genop}). The term $\prob\left\{\adt(R) | \varepsilon_0 = 0\right\}$ is obtained in a similar fashion.

\vspace{-2mm}
\subsection{Proof of theorem \ref{teo:opubound}} \label{proof:opubound}
Analogously to the proof of theorem \ref{teo:opexp} we can show that:
\begin{gather}
\prob\left\{\adt(R) | \varepsilon_0 = 0\right\} = 1 -\mathcal{L}_{I_d}\left(T/l_{sd}\right),\\
\prob\left\{\adf(R) |r,\varepsilon_0 = 1\right\} = 1 -\mathcal{L}_{I_r}\left(T/l_{sr}\right),
\end{gather}
\vspace{-23pt}
\begin{multline}
\prob\left\{\bdf(R)|r,\varepsilon_0 = 1\right\} = \\1  - \frac{D^\alpha \mathcal{L}_{I_d}\left(T/l_{rd}\right) - ||r-d||^\alpha \mathcal{L}_{I_d}\left(T/l_{sd}\right)}{D^\alpha - ||r-d||^\alpha}. \label{eq:pbdfcomplex}
\end{multline}
Using (\ref{eq:laptrans}) from lemma \ref{lemma:Lapcase2} we can evaluate all the LTs. Taking expectation with respect to the relay position $r$ we obtain the general expression. To simplify the expectation of $\prob\left\{\bdf(R)|r,\varepsilon_0 = 1\right\}$ we use (\ref{eq:laptrans}) to evaluate the LT and setting $u = ||r-d||/D$ we write:
\begin{multline}
	\frac{D^\alpha \mathcal{L}_{I_d} \left(T/l_{rd}\right)}{D^\alpha-\|r-d\|^\alpha}- \frac{||r-d||^\alpha \mathcal{L}_{I_d} \left(T/l_{sd}\right)}{D^\alpha-\|r-d\|^\alpha}=\\
	 e^{-\lambda_s \Delta(p_r) D^2} \left(\frac{e^{\lambda_s \Delta(p_r) D^2(1-u^2)}-u^\alpha}{1-u^\alpha}\right).
\end{multline}
Using that as $\lambda_s \Delta D^2(1-u^2) \rightarrow 0$ we have $e^{\lambda_s \Delta D^2(1-u^2)} = 1 + \lambda_s \Delta D^2(1-u^2)+ \Onot((\lambda_s \Delta D^2(1-u^2))^2)$, that for $u>0$:
\begin{equation}
\frac{1-u^2}{1-u^\alpha} \geq 1 + \left(\frac{2}{\alpha}-1\right) u,
\end{equation}
and taking expectation with respect to $u$ we conclude that:
\begin{multline}
\hspace{-5pt}\Ex_r\left[\frac{e^{\lambda_s \Delta(p_r) D^2(1-u^2)}-u^\alpha}{1-u^\alpha}\right] \hspace{-1pt} \geq  \hspace{-1pt}
1 + \lambda_s \Delta(p_r) D^2 \\  \times \left[1+\frac{2-\alpha}{\alpha}\Ex_r\left[u\right] \right] 
 + \Onot\left((\lambda_s \Delta(p_r) D^2)^2 \right).
\end{multline}
To find (\ref{eq:bouNQ20}) start by writing:
 \begin{equation}
 \Ex_r\left[ ||r-d|| \right]  = \int_{\R^2} \frac{1}{2\pi \sigma_{in}^2} ||r-d||e^{-\frac{||r||^2}{2\sigma_{in}^2} } dr. \label{eq:exprd}
 \end{equation}
 Now take $x = r-d$, change to polar coordinates to obtain:
 \begin{align}
 \Ex_r\left[ ||r-d|| \right] &= \int_0^\infty  \left(\frac{u}{\sigma}\right)^2  e^{-\frac{u^2+D^2}{2\sigma^2}} I_0\left(\frac{Du}{\sigma^2}\right) du \\
 &= \sigma Q_{2,0} \left( \frac{D}{\sigma},0\right).
 \end{align}
 In the first step we used the definition of the modified Bessel function. For the actual value of $Q_{2,0}(u,0)$ we use (91) and (60) from \cite{Nuttall1972}.
 
 To find (\ref{eq:ineqErd}) we first prove that:
 \begin{equation}
 ||r-d|| \leq |||d||-||r||| + 2 \min \left(||d||,||r||\right) |\sin\left(\theta/2\right)|, \label{eq:boudistrd}
 \end{equation}
 where $\theta$ is the angle between $r$ and $d$. We decompose $r-d$ as $r-d = u_1 + v_1 = u_2 + v_2$ with:
 \begin{alignat}{2}
 u_1 = r-\frac{||r|| }{||d||}d  \hspace{4mm} &\hspace{4mm} u_2 =  \frac{||d||}{||r||}r-d.
 \end{alignat}
 Then we use that $||u_1|| = 2||r||\sin(\theta/2)$, $||u_2|| = 2||d||\sin(\theta/2)$ and $||v_1|| = ||v_2|| = \left|\left\|r\right\|-\left\|d\right\|\right|$ and the triangle inequality on both decompositions. By taking the expectation on both sides of (\ref{eq:boudistrd}) and solving the integrals we finish the proof.
 
 \vspace{-3mm}

 \subsection{Proofs regarding the concavity of the OP}\label{proof:concavOPrr}
\subsubsection{Proof of theorem \ref{lemma:concavOPrr}} 

We rewrite (\ref{eq:poutrelrand3}) in terms of $\nu(p_r):= \lambda_s \Delta(p_r) D^2 $ to obtain:
\begin{multline}
  \poutm(R) \leq   \left[1-\frac{\alpha (\nu(p_r)-\nu(0))}{2\nu(0)}\right]\left[ 1- e^{- \nu(p_r)}\right] \\ +\frac{\alpha (\nu(p_r)-\nu(0))}{2\nu(0)}\left\{1 +\frac{2\nu(p_r)}{\lambda_{in} \phi_0 D^2 + 2\nu(p_r)}\right.\\ \left.- e^{-\nu(p_r)}  \left[ 1+ \nu(p_r) \left( 1 + \frac{2-\alpha}{\alpha D} \Ex_r \left[||r-d||\right] \right) \right]\right\}. \label{eq:concavOPrr1}
\end{multline}
Since $\nu(p_r)$ is linear in $p_r$ we can analyze the concavity of the OP with respect to $\nu(p_r)$ instead of $p_r$. We do this by studying when the second derivative of the OP upper bound (\ref{eq:concavOPrr1}) with respect to $\nu$ is negative. After differentiating twice with respect to $\nu$ and rearranging the terms we obtain:
\begin{multline}
\frac{d^2\poutm}{d\nu^2} \leq \left\{2\alpha \lambda_{in} \phi_0 D^2e^{\nu(p_r)}\frac{2\nu(0)+ \lambda_{in} \phi_0  D^2}{(2\nu(p_r)+\lambda_{in}\phi_0 D^2)^3} \right.\\ \left.
+\frac{\alpha}{2}\left(1-\frac{2}{\alpha}\right)p_c(\nu) \frac{\Ex_r[||r-d||]}{D} \right.  \\-\left. \left(\frac{\alpha}{2}p_c(\nu)+\nu(0)\right) \phantom{\frac{1}{1}} \hspace{-10pt} \right\}\frac{e^{-\nu(p_r)}}{\nu(0)} , \label{eq:opuboundiff} 
\end{multline}
with:
\begin{equation}
p_c(\nu) = \nu^2 - (\nu(0) + 4) \nu + 2(1+\nu(0)).
\end{equation}
We study the derivative in the interval $0 \leq p_r \leq 1$ which maps to the interval $\nu(0)=\lambda_s \delta D^2 \leq \nu \leq \nu(1) = \lambda_s \delta D^2 (1+2/\alpha)$. Using standard arguments it is straightforward to show that for $\alpha > 2$ we have $p_c(\nu) > 0 $ in this interval whenever:
\begin{equation}
\nu(0)=\lambda_s \delta D^2 \leq \frac{3 - \sqrt{5}}{2} \approx 0.38. \label{eq:conconcavrr1}
\end{equation}
In addition, using that $\nu(p_r) \geq \nu(0) = \lambda_s \delta D^2 > 0$ we bound:
\begin{equation}
\lambda_{in} \phi_0 D^2\frac{2\nu(0)+ \lambda_{in} \phi_0  D^2}{(2\nu(p_r)+\lambda_{in}\phi_0 D^2)^3} \leq  \frac{1}{\lambda_{in} \phi_0  D^2},
\end{equation}
in (\ref{eq:opuboundiff}), to obtain:
\begin{multline}
\frac{d^2\poutm}{d\nu^2} \leq \frac{e^{-\nu(p_r)}}{\nu(0)} \left\{\frac{4\pi \alpha e^{\nu(p_r)}\sigma_{in}^2}{\phi_0  D^2} \right.\\ \left.
+\left(\frac{\alpha}{2}-1\right)p_c(\nu) \frac{\Ex_r[||r-d||]}{D} - \left(\frac{\alpha}{2}p_c(\nu)+\nu(0)\right)\right\},
\label{eq:opuboundiff2} \end{multline}
where we have used also that $\sigma_{in} = (2\pi \lambda_{in})^{-1}$. It is clear that the first term in (\ref{eq:opuboundiff}) is positive and that under (\ref{eq:conconcavrr1}) the second term also is, and the third one is negative. Notice also that for each $\phi_0$, $D$, $\alpha >2$ as $\sigma_{in} \rightarrow 0$, the first term goes to zero and $\Ex_r[||r-d||]/D \rightarrow 1$. Now using standard continuity arguments it is straightforward to show that for each $\phi_0$, $D$, $\alpha >2$ and $\nu(0) = \lambda_s \delta D^2$ satisfying (\ref{eq:conconcavrr1}) if $\sigma_{in}$ is small enough then the third negative term will be greater that the other two, and hence the second derivative will become negative.

\subsubsection{Proof of corollary \ref{coro:concavcond}} 
In order to find the value of $\sigma_c$ we should find the smallest root of the second derivative of $\poutm$. Since this cannot be done in closed form, we can find an approximate condition for concavity by finding the smallest root in $\sigma_{in}$ of the upper bound (\ref{eq:opuboundiff2}) of the second derivative of $\poutm$.
For the high reliability regime we can further approximate $\nu(0) = \lambda_s \delta D^2 \approx 0$ which leads to (\ref{eq:concavcond}). To obtain condition (\ref{eq:condconvclosed}) we upper bound $\Ex[||r-d||]$ using (\ref{eq:ineqErd}) in (\ref{eq:concavcond}), to obtain:

\begin{equation}
2 \alpha s^2 + (\alpha -2 )\gamma(s,\phi_0) s - 2 = 0.\label{eq:approxcondconv}
\end{equation}

This equation cannot be solved in closed form either due to the presence of the function $\gamma$, but it can be shown that an absolute upper bound the smallest root is obtained from setting $\alpha = 2$ in the equation. In that case the equation is independent of $\gamma$ and can be solved in closed form to obtain the condition $s \leq 1/\sqrt{2}$. In can be shown that setting $s = 1/\sqrt{2}$ in $\gamma(s,\phi_0)$ and solving (\ref{eq:approxcondconv}) yields a lower bound on the smallest root for each value of $\alpha$. Thus, (\ref{eq:approxcondconv}) becomes a second degree polynomial in $s$ which can be solved in closed form to obtain (\ref{eq:condconvclosed}).

\subsection{Proof of theorem \ref{lemma:optOPrr}} \label{proof:optOPrr}
In theorem \ref{lemma:concavOPrr} we showed that for each network setup such that (\ref{eq:conconcavrr1}) holds there is an interval $\sigma_{in} \leq \sigma_c$ in which the OP upper bound is concave in $p_r$. Now we show that under this condition, there is an interval in which $p_r = 1$ is optimal by finding conditions such that $\poutm(p_r = 1)  - \poutm(p_r=0) \leq 0$. Setting $p_r = 0$ and $p_r=1$ we can write:
\begin{multline}
\hspace{-12pt}\poutm(p_r = 1) - \poutm(p_r = 0) \leq e^{-\lambda_s \delta D^2} + \frac{2 \lambda_s \Delta}{2 \lambda_s \Delta+ \phi_0 \lambda_{in}} \\ - \left[1 + \lambda_s \Delta D^2 
\left(1 + \frac{2-\alpha}{\alpha D}\Ex_r\left[||r-d||\right]\right)\right] e^{-\lambda_s \Delta D^2}. 
\end{multline}
where we take $\Delta \equiv \Delta(1) = \delta\left(1+\frac{2}{\alpha}\right)$.
Now we upper bound:
\begin{equation}
\frac{2 \lambda_s \Delta}{2 \lambda_s \Delta+ \phi_0 \lambda_{in}} \leq \frac{2 \lambda_s \Delta}{\phi_0 \lambda_{in}} = \frac{4 \pi \lambda_s \Delta \sigma_{in}^2}{\phi_0}
\end{equation}
and:
\vspace{-3mm}
\begin{multline}
e^{-\lambda_s \delta D^2} - e^{-\lambda_s \Delta D^2} \left(1+\lambda_s \Delta D^2\right) \leq  \\
\lambda_s \delta D^2 e^{-\lambda_s \Delta D^2} \left(\frac{4}{\alpha^2}\lambda_s\delta D^2-1\right)
\end{multline}
which is valid when (\ref{eq:conconcavrr1}) is met since then $e^{\nu(0)} \leq 1+{\nu(0)}+{\nu^2(0)}$. With this we obtain:
\vspace{-5mm}

\begin{multline}
\poutm(p_r = 1) - \poutm(p_r = 0) \leq \lambda_s \delta D^2 e^{-\lambda_s \Delta D^2} \\ \left\{ \left[1+\frac{2}{\alpha}\right] \left[\frac{4 \pi \sigma_{in}^2}{\phi_0D^2} e^{\lambda_s \Delta D^2} + \left(1-\frac{2}{\alpha}\right)\frac{\Ex_r[||r-d||]}{D}\right] \right. \\ \left.
+ \frac{4\lambda_s \delta D^2}{\alpha^2} -1\right\}.
\label{eq:poutpr1}
\end{multline}

Continuity arguments similar to those of theorem \ref{lemma:concavOPrr} prove that if $\sigma_{in}$ is small enough then the right side of (\ref{eq:poutpr1}) will be negative and $p_r=1$ will be optimal. To find and estimate for the maximum value of $\sigma_{in}$  we can find the roots of the right side of this expression, focusing on the terms between brackets.  In the high reliability regime the term $\lambda_s \delta D^2$ will be small (as shown at the end of Section II) so an approximate condition for concavity can be obtained by letting $\lambda_s \delta D^2 \rightarrow 0$, which leads to (\ref{eq:condpr1opt}).
The proof of the simpler condition (\ref{eq:pr1opt2}) is obtained following the same arguments as in the proof of corollary \ref{coro:concavcond}, except that in this case we can establish the condition $s \leq 1/2$.

\subsection{Proof of theorem \ref{teo:opgain}} 
\label{proof:opgain}
When $p_r = 1$ is optimal we can compute the gains starting from (\ref{eq:poutpr1}), valid under (\ref{eq:conconcavrr1}), and noting that $\poutm(p_r = 0) = \poutdt$. Rearranging the terms we obtain:
\begin{multline}
\frac{\poutm}{\poutdt} \leq 1 + \frac{\lambda_s \delta D^2}{\poutdt}\left\{\left[\left(1-\frac{2}{\alpha}\right) \frac{\Ex_r[||r-d||]}{D} e^{-\lambda_s \Delta D^2} \right.\right.\\ \left. \left.
+ \frac{4\pi\sigma_{in}^2}{\phi_0 D^2}\right] \hspace{-3pt}
\times  \hspace{-3pt} \left(1+\frac{2}{\alpha}\right) + e^{-\lambda_s \Delta D^2} \left(\frac{4\lambda_s\delta D^2}{\alpha^2}-1\right)\right\}.
\end{multline}
By noting that when $p_r = 1$ is optimal the term between brackets in the previous expression will be negative, we can upper bound this by removing the term $(\lambda_s \delta D^2)/\poutdt$ outside the brackets. To simplify the expression for the high reliability regime, we can take the approximation $\lambda_s \delta D^2 \approx 0$. Finally, when $p_r = 0$ the performance will be the same as DT so the gain will be one.

\bibliographystyle{IEEEtran}
\bibliography{IEEEabrv,relaygeo}

\begin{IEEEbiography}[{\includegraphics[width=1in,height=1.25in,clip,keepaspectratio]{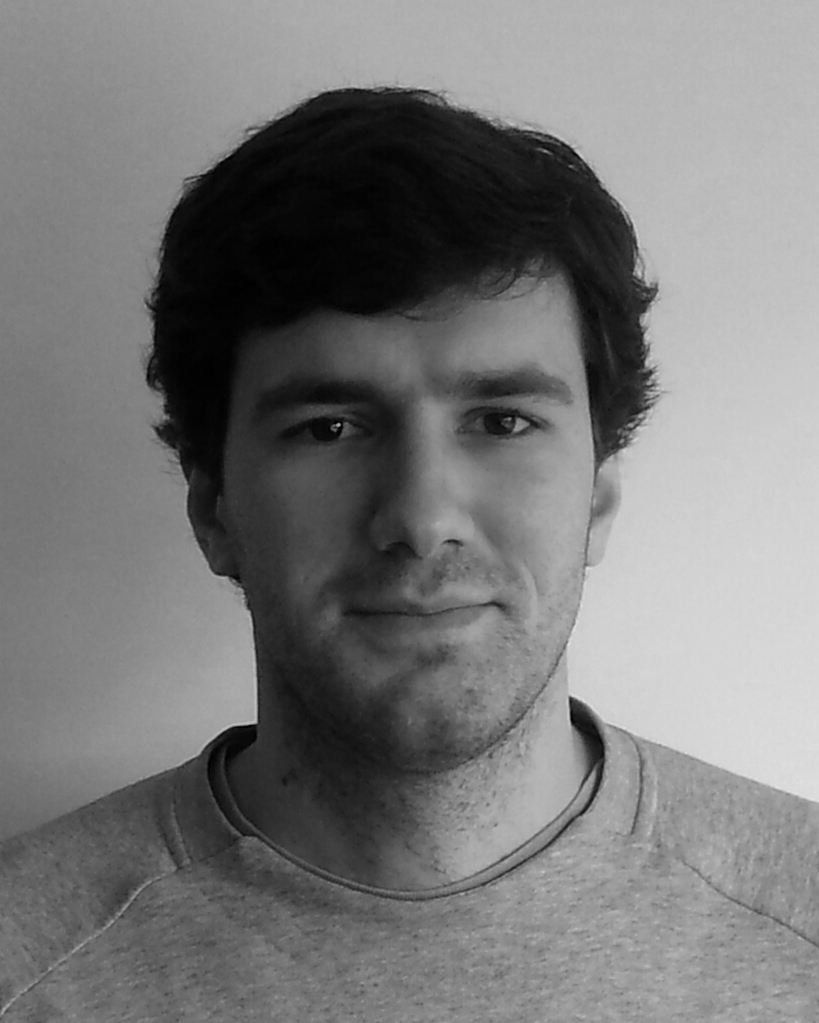}}]{Andrés Altieri}
(S'12) received the B.Sc. and M.Sc degrees (with honors) in Electrical Engineering from the University of Buenos Aires (Argentina) in 2009. He is currently pursuing a joint PhD degree between the University of Buenos Aires and the École Supérieure d'Électricité (SUPELEC, France).

His research interests include cooperative wireless networks, information theory and stochastic geometry.
\end{IEEEbiography}
\vspace*{-8mm}

\begin{IEEEbiography}[{\includegraphics[width=1in,height=1.25in,clip,keepaspectratio]{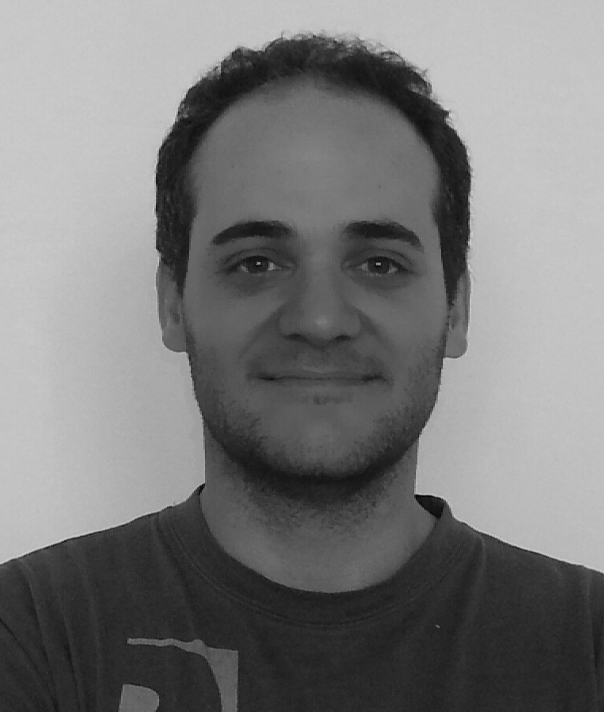}}]{Leonardo Rey Vega} received the M.Sc (with honors) and PhD  degrees in Electrical Engineering from the University of Buenos Aires (Argentina) in 2004 and 2010, respectively. 

In 2007 and 2008 he was invited at the INRS-EMT in Montreal, Canada and in the first semester 2012 he was a visitor at the Department of Telecommunications at SUPELEC, France. He is currently an Assistant Professor at the University of Buenos Aires.

 Dr. Rey Vega's research interests include information theory, cooperative communications, and statistical signal processing.
\end{IEEEbiography}
\vspace*{-8mm}
\begin{IEEEbiography}[{\includegraphics[width=1in,height=1.25in,clip,keepaspectratio]{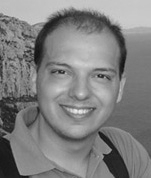}}]{Pablo Piantanida}
(S'04-M'08) received the B.Sc. and M.Sc degrees (with honors) in Electrical Engineering from the University of Buenos Aires (Argentina) in 2003, and the Ph.D. from the Paris-Sud University (France) in 2007. 

In 2006, he was with the Department of Communications and Radio-Frequency Engineering at Vienna University of Technology (Austria). In October 2007 he joined the Department of Telecommunications, SUPELEC, as an Assistant Professor in network information theory. 

Dr Piantanida's research interests include multi-terminal information theory, Shannon theory, cooperative communications, physical-layer security and coding theory for wireless applications. 
\end{IEEEbiography}

\begin{IEEEbiography}[{\includegraphics[width=1in,height=1.25in,clip,keepaspectratio]{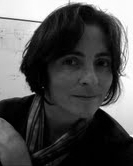}}]{Cecilia Galarza} received the Ingeniera Electrónica degree from the
University of Buenos Aires, Buenos Aires, Argentina, in 1990, and the
M.S. and Ph.D degrees in electrical engineering and computer science
from the University of Michigan, Ann Arbor, U.S.A in 1995 and 1998
respectively.

She is currently an Associate Professor in the School of Engineering
at the University of Buenos Aires (FIUBA) in Buenos Aires, Argentina.
Before joining the FIUBA she worked at Voyan Technologies, Santa
Clara, CA.

Prof. Galarza is a researcher at the National Council for Scientific
and Technological Research  (CONICET) in Argentina. Her research
interests include wireless networks, process detection on sensor
networks, and sensor management
\end{IEEEbiography}

\end{document}